\documentclass[twocolumn,traditabstract]{aa}
\usepackage{amssymb}
\usepackage{mathptmx}
\usepackage{natbib}
\bibpunct{(}{)}{;}{a}{}{,}
\usepackage[]{graphicx}
\usepackage{xcolor}
\usepackage{lscape}
\usepackage{txfonts}
\usepackage{verbatim}
\usepackage{url}
\defcitealias{canameras15}{C15}
\defcitealias{nesvadba19}{N19}
\defcitealias{canameras17b}{C17a}
\defcitealias{canameras18b}{C18}

\begin{document}

\title{\textit{Planck}'s Dusty GEMS. VIII. Dense-gas reservoirs in the most active dusty starbursts at z$\sim$3
  \thanks{Based on IRAM data obtained with programs S15CH and 108--14, and on ALMA data from program 2015.1.01518.}}

\author{R. Ca\~nameras\inst{1}\and N. P. H. Nesvadba\inst{2}\and R. Kneissl\inst{3,4}\and S. K\"onig\inst{5}\and C. Yang\inst{3}\and A. Beelen\inst{6}\and R. Hill\inst{8}\and E. Le Floc'h\inst{7}\and D. Scott\inst{8}}

\institute{
Max-Planck-Institut f\"ur Astrophysik, Karl-Schwarzschild-Str. 1, 85748 Garching, Germany \\
{\tt e-mail: rcanameras@mpa-garching.mpg.de}
\and
Universit\'e C\^ote d'Azur, Observatoire de la C\^ote d'Azur, CNRS, Laboratoire Lagrange, France
\and
European Southern Observatory, ESO Vitacura, Alonso de Cordova 3107, Vitacura, Casilla 19001, Santiago, Chile
\and
Atacama Large Millimeter/submillimeter Array, ALMA Santiago Central Offices, Alonso de Cordova 3107, Vitacura, Casilla 763-0355,
Santiago, Chile
\and
Department of Space, Earth and Environment, Onsala Space Observatory, Chalmers University of Technology, 43992 Onsala, Sweden
\and
Institut d'Astrophysique Spatiale, CNRS, Univ. Paris-Sud, Universit\'e Paris-Saclay, B\^at. 121, 91405 Orsay France
\and
Laboratoire AIM, CEA/DSM/IRFU, CNRS, Universit\'e Paris-Diderot, B\^at. 709, 91191 Gif-sur-Yvette, France
\and
Department of Physics and Astronomy, University of British Columbia, 6224 Agricultural Road, Vancouver 6658, British Columbia,
Canada
}

\titlerunning{\textit{Planck}'s Dusty GEMS. VIII. Dense-gas tracers}

\authorrunning{R. Ca\~nameras et al.} \date{Received / Accepted}

\abstract{We present ALMA, NOEMA, and IRAM-30m/EMIR observations of the high-density tracer molecules HCN, HCO$^+$, and HNC in three of
  the brightest lensed dusty star-forming galaxies at $z \simeq 3$--3.5, part of the {\it Planck}'s Dusty Gravitationally Enhanced
    subMillimetre Sources (GEMS), with the aim of probing the gas reservoirs closely associated with their exceptional levels of star
  formation. We obtained robust detections of ten emission lines between $J_{\rm up}=4$ and 6, as well as several additional upper flux
  limits. In PLCK\_G244.8+54.9, the brightest source at $z=3.0$, the HNC(5--4) line emission at 0.1\arcsec\ resolution, together with
  other spatially-integrated line profiles, suggests comparable distributions of dense and more diffuse gas reservoirs, at least over the
  most strongly magnified regions. This rules out any major effect from differential lensing. This line is blended with CN(4--3) and in
  this source, we measure a HNC(5--4)/CN(4--3) flux ratio of 1.76$\pm$0.86. Dense-gas line profiles generally match those of mid-$J$ CO
  lines, except in PLCK\_G145.2+50.8, which also has dense-gas line fluxes that are relatively lower, perhaps due to fewer dense cores
  and more segregated dense and diffuse gas phases in this source. The ${\rm HCO^+/HCN \gtrsim 1}$ and ${\rm HNC/HCN \sim 1}$ line ratios
  in our sample are similar to those of nearby ultraluminous infrared galaxies (ULIRGs) and consistent with photon-dominated regions
  without any indication of important mechanical heating or active galactic nuclei (AGN) feedback. We characterize the dense-gas excitation
  in PLCK\_G244.8+54.9 using radiative transfer models assuming pure collisional excitation and find that mid-$J$ HCN, HCO$^+$, and HNC
  lines arise from a high-density phase with an H$_{\rm 2}$ density of $n\,{\sim}\,10^5$--$10^6\,{\rm cm}^{-3}$, although important degeneracies
  hinder a determination of the exact conditions. The three GEMS are consistent with extrapolations of dense-gas star-formation laws
  derived in the nearby Universe, adding further evidence that the extreme star-formation rates observed in the most active galaxies
  at $z \sim 3$ are a consequence of their important dense-gas contents. The dense-gas-mass fractions traced by HCN/[CI] and HCO$^+$/[CI]
  line ratios are elevated, but not exceptional as compared to other lensed dusty star-forming galaxies at $z>2$, and they fall near
  the upper envelope of local ULIRGs. Despite the higher overall gas fractions and local gas-mass surface densities observed at high
  redshift, the dense-gas budget of rapidly star-forming galaxies seems to have evolved little between $z \sim 3$ and $z \sim 0$. Our
  results favor constant dense-gas depletion times in these populations, which is in agreement with theoretical models of star formation.}


\maketitle

\section{Introduction}
\label{sec:intro}

Observational studies of the bright emission lines of $^{12}$CO have greatly improved our understanding of the global molecular gas
content and relationship to star formation in galaxies from low \citep[e.g.,][]{kennicutt98} to high redshifts
\citep[e.g.,][]{daddi10a,genzel10,carilli13}. However, most of these lines are optically thick and do not probe the dense cloud cores
that are unstable with regard to gravitational collapse and that directly fuel star formation. Characterizing molecular gas in this
high-density phase is, therefore, fundamental to test star-formation models, as well as to constrain the main (radiative and non-radiative)
feedback mechanisms driving turbulence in pre-stellar cores and slowing down the collapse of the gas. Rotational transitions of HCN,
HCO$^+$, and HNC are usually collisionally excited by H$_{\rm 2}$ in environments with densities above $10^5$--$10^6\,{\rm cm}^{-3}$, about
three orders of magnitude higher than $^{12}$CO for a given $J$ level, and have been extensively used to study the dense cores of molecular
clouds in the Milky Way \citep[e.g.,][]{helfer97}. Despite their low brightness (with fluxes about one order of magnitude lower than for
$^{12}$CO), these emission lines also hold a great potential for tightly constraining the dense-gas phase, its mass fraction, and the
relation to on-going star formation in external galaxies \citep[e.g.,][]{gao04b,gao04a,garciaburillo12}. Recent hydrodynamical simulations
of star-forming clouds confirm that these molecules are effective and versatile tracers of dense gas for environments with a range of
star-formation rates \citep[see, e.g.,][for HCN]{onus18}. In rare cases, ground-level emission of these molecules has also been detected
at much lower densities \citep[e.g., down to $\simeq 10^3\,{\rm cm}^{-3}$ in the Orion A cloud,][]{kauffmann17}, however, \citet{evans20}
point out that for starburst galaxies, this component should have a minor contribution to the overall emission line fluxes.

Observational studies of the dense-gas reservoirs in low-redshift star-forming galaxies started about three decades ago
\citep[e.g.,][]{nguyen92}, followed by extended surveys of HCN line emission in luminous and ultra-luminous infrared galaxies (ULIRGs)
and normal spirals \citep{gao04b,gao04a}. Since then, HCN and HCO$^+$ detections have been extensively used to characterize the dense
interstellar medium (ISM) of local (U)LIRGs \citep[e.g.,][]{graciacarpio08,juneau09,garciaburillo12}, including diagnostics for the
physical conditions of the gas from line excitation modeling \citep[see, e.g.,][for NGC~6240 and Arp~193]{papadopoulos14} and
spatially-resolved studies with the NOrthern Extended Millimeter Array (NOEMA) and Atacama Large Millimeter/submillimeter Array (ALMA)
interferometers \citep{chen17,konig18,imanishi19}. The reliability of each molecule as a proxy of the overall dense-gas content was
discussed in \citet{papadopoulos07}. Recently, valuable constraints on the dense-gas components of galactic outflows have also allowed
us to peer into the embedded star-formation activity and its role as a positive or negative feedback mechanism
\citep[e.g.,][]{walter17,michiyama18,aladro18}. Combining sub-arcsec resolution HCN and HCO$^+$ maps with traditional CO lines provides
particularly detailed pictures of complex molecular gas flows and triggering mechanisms, as demonstrated in the local barred Seyfert
galaxy NGC~1068 \citep{garciaburillo14} and the Medusa \citep{konig18}.

Following expectations that HCN is emitted from regions with active star formation, early studies of local ULIRGs found a tight linear
correlation between the total IR and HCN(1--0) line luminosities \citep{gao04a} and unveiled the impact of variations in dense-gas
fractions on the global Schmidt-Kennicutt law \citep{kennicutt98} in galaxies. \citet{wu05} demonstrated that this relation holds down
to the small scales of dense Galactic cloud cores, thereby extending its validity over at least seven orders of magnitude in infrared
luminosity \citep[see also][]{wu10,stephens16}. Subsequent studies discussed the behavior of this linear correlation for various high-density
tracer molecules, such as HCO$^+$ \citep[e.g.,][]{papadopoulos07}, for higher $J$ levels \citep[e.g.,][]{graciacarpio08,zhang14}, and
for high-redshift galaxies \citep[e.g.,][]{gao07}. Most of these extragalactic studies have relied on galaxy-integrated measurements while
focusing on central nuclear regions that host most of the dense gas under extreme ISM conditions. \citet{tan18} supplemented these results
with spatially resolved, sub-kpc scale observations of HCN and HCO$^+$ $J=4$--3 in six of the brightest nearby LIRGs. They showed that
lower surface-density regions over the disks are broadly consistent with $L_{\rm FIR}-L'_{\rm dense}$ relations on global scales. Theoretical
models have suggested that the linear slopes observed for dense-gas star-formation laws (lower than for CO(1--0)) can be explained for
transitions with critical densities much higher than the average ISM density \citep{krumholz07b}. However, \citet{zhang14} later obtained
slopes of unity for transitions with very high critical density as well.

These results have been interpreted as evidence that the star-formation activity of galaxies out to high redshift is essentially driven by
the mass fraction of their dense molecular gas, with a roughly constant star-formation efficiency (SFE) per free-fall time in the dense-gas
phase \citep[][]{krumholz07a}. This is consistent with turbulence-regulated star-formation models \citep{krumholz05}. Sub-kpc properties of
local star-forming galaxies show that these dense-gas fractions are themselves strongly correlated with the average molecular gas-mass
surface density of clouds since both quantities are inherited from the overall gas density distribution \citep{gallagher18}.

Furthermore, non-collisional processes can enhance the luminosities of HCN, HCO$^+$, and HNC lines in regions with low dense-gas fractions
and vice versa. For instance, HCN and HCO$^+$ can be excited by IR-pumping \citep{aalto95}, and HCO$^+$ emission drops for a high abundance
of free electrons and is therefore highly sensitive to the ambient cosmic-ray fluxes \citep{papadopoulos07}. Measuring line ratios helps to
identify these processes. The HNC/HCN ratio provides valuable diagnostics of photo-dissociation regions \citep[PDRs,][]{meijerink07},
X-ray-dominated regions (XDRs) and IR-pumping \citep{aalto07}, and shock-dominated regions \citep{loenen08,aladro15}. The HCO$^+$/HCN
line-intensity ratio is also widely used to discriminate between active galactic nuclei- (AGN) and starburst-dominated galaxies, given the
apparent dichotomy between both populations  \citep[e.g.,][]{krips08,imanishi14,imanishi16,izumi16}. Other studies, such as that of
\citet{privon15}, nonetheless find additional dependences on ISM density and the surrounding radiation fields
\citep[e.g., IR-pumping,][]{krips08}, and demonstrate that HCO$^+$/HCN can strongly vary in LIRGs and ULIRGs experiencing both AGN and star
formation feedback within their unresolved central regions. In NGC~1068, the HCO$^+$/HCN ratio is even enhanced by the central AGN
\citep{garciaburillo14}.

The dense ISM components remain poorly characterized at high-redshift due to the difficulty in detecting these faint transitions, and most
efforts have focused on powerful quasi-stellar objects (QSOs) such as the Cloverleaf \citep{wilner95,solomon03,riechers06,riechers10,riechers11},
or on extremely bright sub-millimeter galaxies (SMGs). Early efforts included upper HCN flux limits measured for two bright SMGs by
\citet{greve06}, and HCN detections in two SMGs and two QSOs from \citet{gao07}. Many subsequent studies have benefited from the flux boosts
due to strong lensing magnifications. In the Cosmic Eyelash, \citet{danielson11} obtained a detection of HCN that was combined with other
tracers to probe the ISM conditions. Dense-gas diagnostics were also inferred from HCN, HCO$^+$, and HNC in two H-ATLAS lensed SMGs at $z=1.58$
and $z=1.79$ \citep{oteo17}, from HCN, HCO$^+$, HNC, and $^{13}$CO in five lensed SMGs at $2.5<z<4$ from the South Pole Telescope (SPT) sample
\citep[][]{bethermin18}, and from H$_{\rm 2}$O and H$_{\rm 2}$O$^+$ emission lines \citep[][]{yang16}. In addition, average line properties of
$^{13}$CO, HCN, HNC, HCO$^+$, CN, and H$_{\rm 2}$O were measured in SPT lensed SMGs using a stacking analysis \citep{spilker14}.

\begin{table*}
  \caption{Details of the NOEMA, ALMA, and IRAM-30m/EMIR observations.}
\centering
\tiny
\begin{tabular}{lllllllll}
\hline
\hline
\noalign{\vskip 1pt}
Source & Line(s) & Project ID & Observing dates & Beam size & $\nu_{\rm tuning}$ & $t_{\rm obs}$ & rms & $S/T^*_{\rm a}$ \\
 & & & & & (GHz) & (hr) & (mK) & (Jy/K) \\
\hline
\noalign{\vskip 1pt}
{\it NOEMA} \\
PLCK\_G092.5+42.9 & HCN, HCO$^+$ $J=5$--4 & S15CH & 09/15 & 6.1\arcsec$\times$\,4.0\arcsec & 105.03 & 3.9 & 0.83$^{(*)}$ & \dots \\
PLCK\_G145.2+50.9 & HCN, HCO$^+$, HNC $J=5$--4 & S15CH & 10/15 & 5.6\arcsec$\times$\,4.3\arcsec & 98.25 & 3.9 & 0.75$^{(*)}$ & \dots \\
PLCK\_G244.8+54.9 & HCN, HCO$^+$, HNC $J=5$--4 & S19CU & 02/12/19 & 2.9\arcsec$\times$\,1.2\arcsec & 112.20 & 6.0 & 0.66$^{(*)}$ & \dots \\
\hline 
\noalign{\vskip 1pt}
{\it ALMA} \\
PLCK\_G244.8+54.9 & HCN $J=5$--4 & 2015.1.01518 & 21/07/16 & 0.78\arcsec$\times$\,0.67\arcsec & 110.63 & 1.2 & $\simeq$0.2$^{(*)}$ & \dots \\
 & HNC $J=5$--4 (LR) & 2015.1.01518 & 21/07/16 & 0.78\arcsec$\times$\,0.67\arcsec & 113.12 & 1.2 & $\simeq$0.2$^{(*)}$ & \dots \\
 & HNC $J=5$--4 (HR) & 2015.1.01518 & 23/10/15 & 0.12\arcsec$\times$\,0.08\arcsec & 112.90 & 1.2 & $\simeq$0.8$^{(*)}$ & \dots \\
\hline
\noalign{\vskip 1pt}
{\it IRAM-30m/EMIR} \\
PLCK\_G092.5+42.9 & HNC $J=4$--3 & 094--13 & 10/04/13 & 28\arcsec$\times$\,28\arcsec & 89.40 & 2.4 & 1.03 & 5.9 \\
 & HCN, HCO$^+$ $J=5$--4 & 108--14 & 20, 21, 22, 23/06 \& 17/09/14 & 23\arcsec$\times$\,23\arcsec & 104.70 & 10.9 & 0.33 & 5.9 \\
 & HNC $J=5$--4 & 108--14 & 20, 21, 22, 23/06 \& 17/09/14 & 23\arcsec$\times$\,23\arcsec & 104.70 & 10.9 & 0.42 & 6.0 \\
 & HCN, HCO$^+$ $J=7$--6 & 108--14 & 20, 21, 22, 23/06/14 & 17\arcsec$\times$\,17\arcsec & 147.00 & 15.1 & 0.34 & 6.2 \\
 & HNC $J=7$--6 & 108--14 & 20, 21, 22, 23/06/14 & 17\arcsec$\times$\,17\arcsec & 147.00 & 13.5 & 0.38 & 6.3 \\
PLCK\_G145.2+50.9 & HCN, HCO$^+$ $J=5$--4 & 223--13 \& 108--14 & 18/04 \& 19, 20/06 \& 18/09/14 & 25\arcsec$\times$\,25\arcsec & 97.80 & 13.2 & 0.31 & 5.9 \\
 & HNC $J=5$--4 & 108--14 & 19, 20/06 \& 18/09/14 & 25\arcsec$\times$\,25\arcsec & 97.80 & 9.2 & 0.36 & 5.9 \\
 & HCN, HCO$^+$, HNC $J=7$--6 & 108--14 & 19, 20/06 \& 18/09/14 & 18\arcsec$\times$\,18\arcsec & 137.50 & 11.2 & 0.35 & 6.1 \\
PLCK\_G244.8+54.9 & HCN, HCO$^+$, HNC $J=4$--3 & 108--14 & 17, 18, 19/06/14 & 27\arcsec$\times$\,27\arcsec & 89.00 & 16.8 & 0.21 & 5.9 \\
 & HNC $J=5$--4 & 094--13 \& 223--13 & 06/06/13 \& 30, 31/01/14 & 21\arcsec$\times$\,21\arcsec & 114.80 & 3.2 & 1.78 & 6.0 \\
 & HCN, HCO$^+$, HNC $J=6$--5 & 108--14 & 17, 18, 19/06/14 & 18\arcsec$\times$\,18\arcsec & 134.00 & 16.2 & 0.24 & 6.1 \\[+0.5em]
PLCK\_G045.1+61.1 & HCN, HCO$^+$ $J=5$--4 & 094--13 \& 223--13 & 07, 10/06/13 \& 21, 22/04/14 & 24\arcsec$\times$\,24\arcsec & 101.90 & 2.2 & 1.00 & 5.9 \\
PLCK\_G113.7+61.0 & HCN $J=4$--3 & 094--13 & 09/04/13 & 24\arcsec$\times$\,24\arcsec & 101.90 & 4.8 & 0.95 & 5.9 \\
 & HCO$^+$, HNC $J=4$--3 & 094--13 & 09/04/13 & 24\arcsec$\times$\,24\arcsec & 101.90 & 2.4 & 1.12 & 5.9 \\
PLCK\_G138.6+62.0 & HCN, HCO$^+$ $J=4$--3 & 094--13 & 09/04/13 & 24\arcsec$\times$\,24\arcsec & 101.90 & 3.2 & 1.23 & 5.9 \\
PLCK\_G165.7+67.0 & HCN, HCO$^+$ $J=6$--5 & 223--13 & 31/01, 01, 03, 04/02/14 & 15\arcsec$\times$\,15\arcsec & 164.57 & 2.5 & 0.53 & 6.5 \\
 & HNC $J=6$--5 & 223--13 & 31/01, 01, 03, 04/02/14 & 15\arcsec$\times$\,15\arcsec & 164.57 & 5.2 & 0.43 & 6.5 \\
PLCK\_G200.6+46.1 & HCN, HCO$^+$ $J=4$--3 & 094--13 & 08, 10/06/13 & 28\arcsec$\times$\,28\arcsec & 89.40 & 3.0 & 0.63 & 5.9 \\
\hline
\end{tabular}
\tablefoot{Here, $\nu_{\rm tuning}$ is the frequency setup of the corresponding receiver and $t_{\rm obs}$ the total integration time on source.
  For EMIR spectra, total exposure times $t_{\rm obs}$ only include scans selected for the optimal reduction. The HNC(5--4) line in
  PLCK\_G244.8+54.9 was observed both at lower (LR) and higher (HR) spatial resolution with ALMA (see details in the text). The rms noise
  levels are measured on line-free spectral channels of source-integrated spectra and given for channels binned to 50--$55\,{\rm km}\,{\rm s}^{-1}$,
  $100\,{\rm km}\,{\rm s}^{-1}$, and 35--$40\,{\rm km}\,{\rm s}^{-1}$ for IRAM-30m/EMIR, NOEMA, and ALMA spectra, respectively. Telescope
  efficiencies $S/T^*_{\rm a}$ listed for EMIR observations are extrapolated from the calibration tables and used to convert the line fluxes
  to ${\rm Jy}\,{\rm km}\,{\rm s}^{-1}$. The rms values marked with an asterisk are given in mJy rather than mK.} 
\label{tab:obslog}
\end{table*}

Pursuing dense-gas studies of dusty star-forming galaxies (DSFGs) at high-redshift, without dominant AGN heating, is thus vital to improving
our understanding of the mode of star formation governing the major phase of galaxy growth. Characterizing the physical properties and SFE
of their dense-gas reservoirs addresses the ongoing debate on whether they are the analogues of nearby intensely star-forming galaxies,
despite their much higher gas-mass and star-formation surface densities, or whether there is a distinct mode of star formation at high
redshift \citep{daddi10a,genzel10}.

In this paper, we study the dense-gas components of {\it Planck}'s Dusty Gravitationally Enhanced subMillimetre Sources (GEMS) -- a sample of
11 of the brightest gravitationally lensed DSFGs on the extragalactic sky, with (at most) minor AGN contributions to the far-IR (FIR) luminosity
\citep{canameras15}. Our extensive CO and [\ion{C}{i}] emission-line survey with the Eight MIxer Receiver (EMIR) on the IRAM 30-m telescope from
\citet[][hereafter \citetalias{canameras18b}]{canameras18b} and \citet[][hereafter \citetalias{nesvadba19}]{nesvadba19} revealed a range of gas
excitations, with CO spectral line energy distributions (SLEDs) peaking in the range of $J_{\rm up}=4$--7, and with a plateau up to $J_{\rm up}=10$,
and demonstrated that the GEMS contain vast amounts of warm, dense molecular gas. Detailed excitation models have also shown evidence for an
additional, fainter, and lower density gas phase in some sources. Here, we use observations of mid-$J$ HCN, HCO$^+$, and HNC lines in eight GEMS
with IRAM-30m/EMIR, NOEMA, and ALMA to probe their densest ISM phase, which is more representative of cloud cores than previous CO line
detections. In particular, the primary focus of this paper is on the three brightest GEMS, namely, PLCK\_G092.5+42.9, PLCK\_G145.2+50.9, and
PLCK\_G244.8+54.9 (``the Ruby''), the latter being a maximal starburst\footnote{Forming most of its stellar mass within one to a few dynamical
  times, with star-formation intensities approaching the highest values expected for a vertically-supported, self-gravitating gas disk
  \citep[e.g.,][]{thompson05}.} at $z=3$ with local star-formation intensities of up to $2000\,{\rm M}_{\odot}\,{\rm yr}^{-1}\,{\rm kpc}^{-2}$,
but rather moderate SFEs of 1--10\%\footnote{Defined as the fraction of the available gas mass converted into stars per free-fall time.} akin
to giant molecular clouds in the Milky Way \citep{canameras17b}. Our high-resolution PdBI, SMA, and ALMA $^{12}$CO and dust continuum
interferometry indicate that PLCK\_G092.5+42.9 and PLCK\_G145.2+50.9 form giant arcs \citep{canameras15}, while PLCK\_G244.8+54.9 forms a
partial Einstein ring \citep{canameras17a}. These three GEMS are aligned with massive foreground galaxies at intermediate redshifts and our
detailed strong-lensing models with {\sc Lenstool} \citep{jullo07} show that they are strongly magnified by factors of $\mu \simeq 10$--22
\citepalias[][Ca\~nameras et al., in prep.]{canameras18b}.

\begin{figure*}
  \centering
  \includegraphics[height=0.26\textwidth]{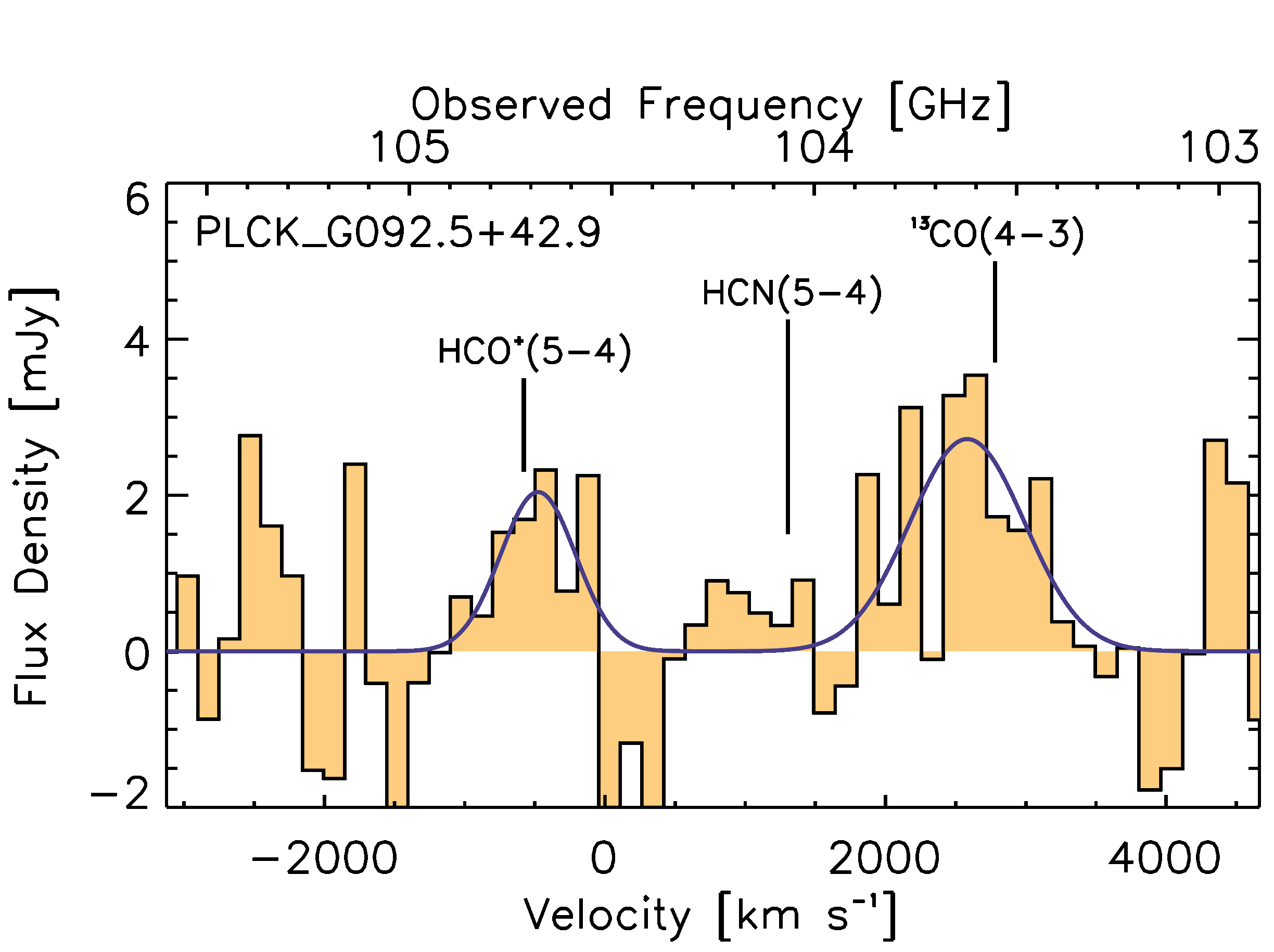}
  \includegraphics[height=0.26\textwidth]{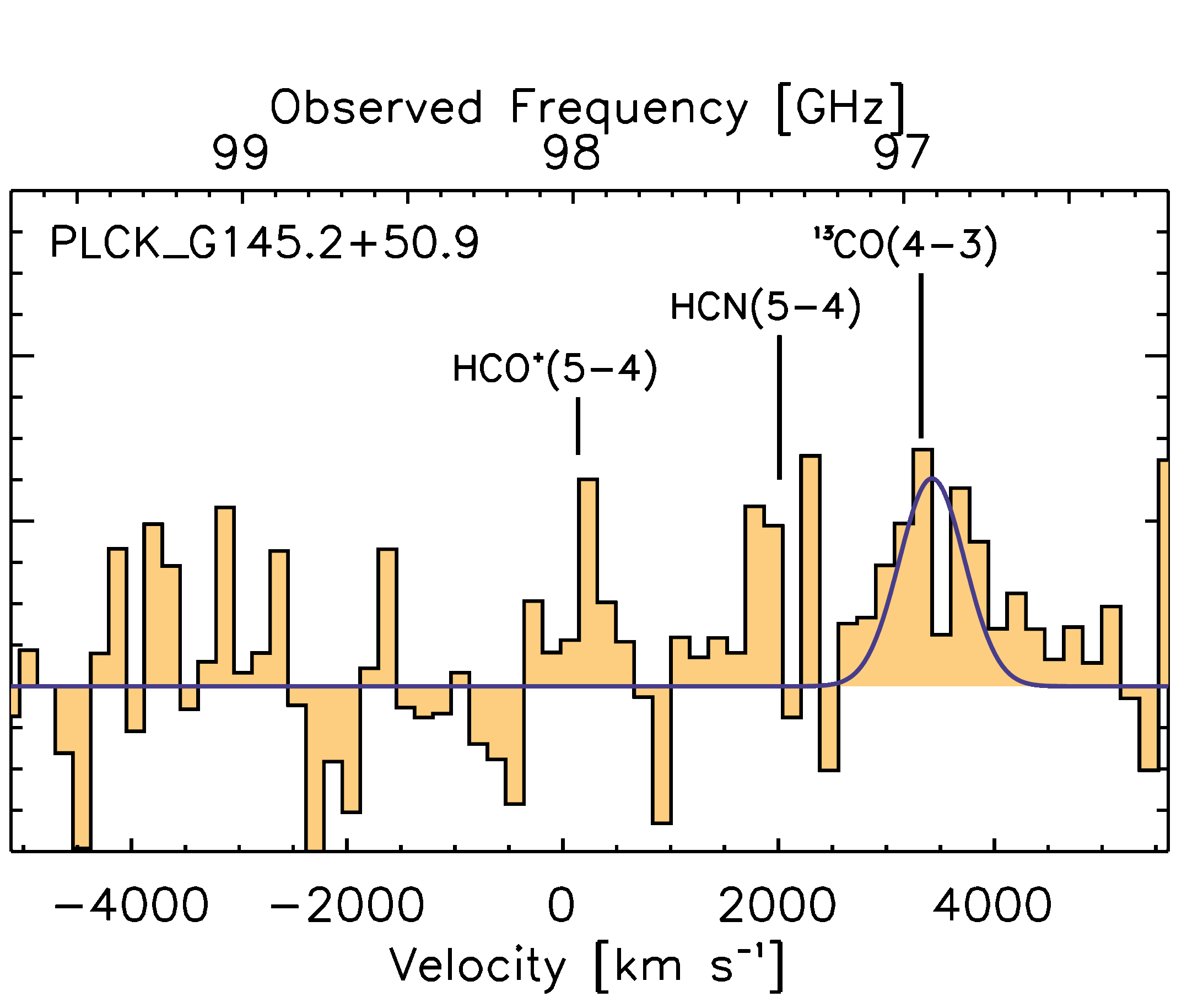}
  \includegraphics[height=0.27\textwidth]{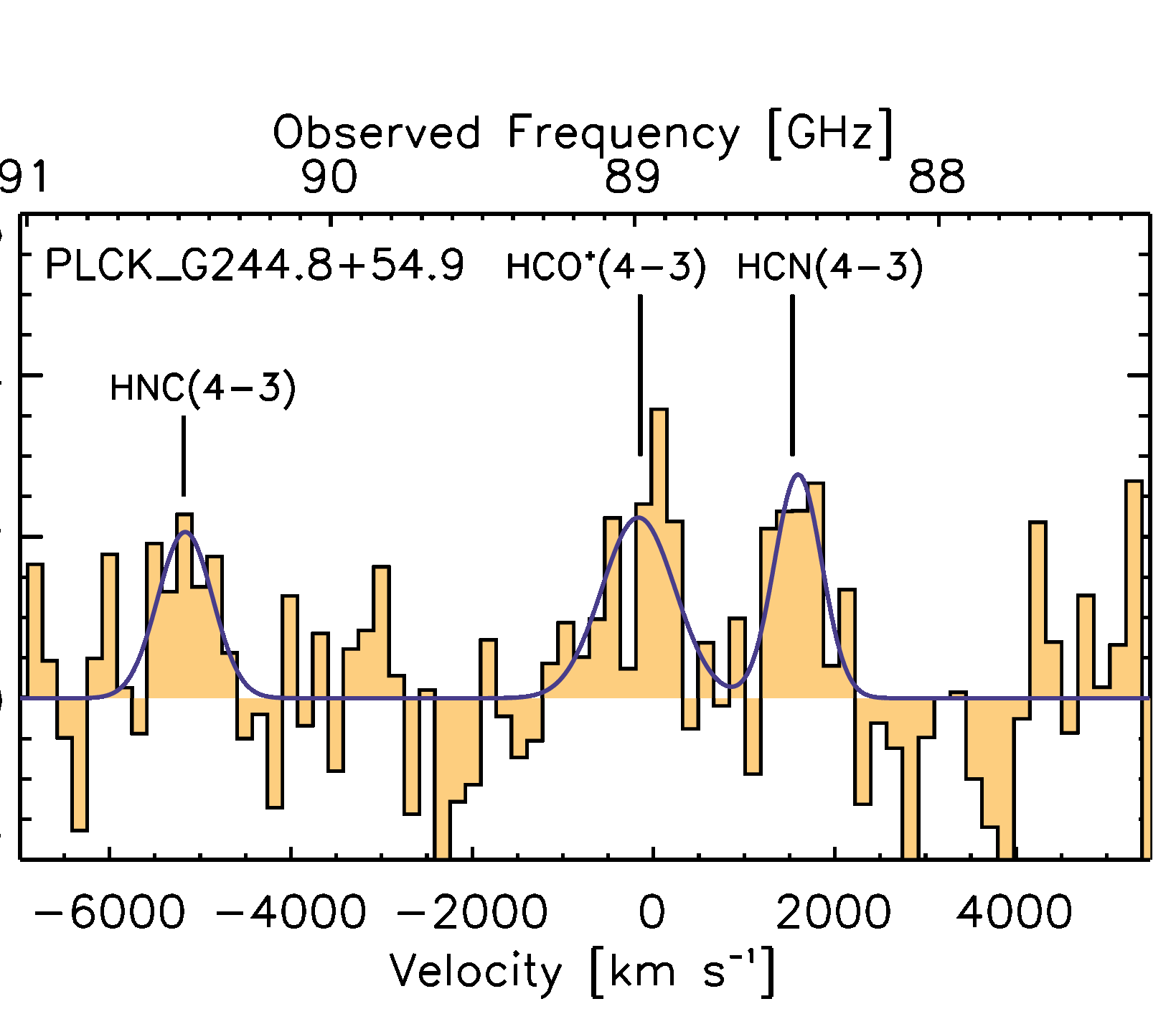}
  \caption{Spatially integrated EMIR spectra of emission lines from high-density tracer molecules observable using a single setup in the 3-mm
    band. The spectra were continuum-subtracted, binned to 50--$55\,{\rm km}\,{\rm s}^{-1}$ resolution, and lines detected at ${>}\,3\,\sigma$
    were fitted with single Gaussian functions using the {\tt CLASS} software package from {\tt GILDAS} (blue curves). Additional tests were
    performed to confirm that these detections are real (see details in Sect.~\ref{ssec:emirobs}).}
  \label{fig:emir}
\end{figure*}

The outline of the paper is as follows. In Sect.~\ref{sec:specobs}, we summarize the observations and describe the data reduction procedures.
In Sect.~\ref{sec:lineprop}, we present the line properties. In Sect.~\ref{sec:denseprop}, we discuss the diagnostics we use for the dense-gas
reservoirs from line ratios and we then analyze the dense-gas star-formation law in Sect.~\ref{sec:sflaw}. The detections and upper limits of
$^{13}$CO from contiguous spectral windows will be presented in a future paper. In addition to $v=0$ lines of HCN, HCO$^+$, and HNC, our observing
setups generally cover the $v=1$ line of vibrationally-excited HCN, but the latter is not seen and also not expected to contribute significantly
in non-AGN sources and, thus, we ignore it in our analysis. We adopt the best-fit flat $\Lambda$CDM cosmology from \citet{planck16}, with
$H_0=67.81\,{\rm km}\,{\rm s}^{-1}\,{\rm Mpc}^{-1}$, $\Omega_{\rm M}=0.308$, and $\Omega_\Lambda=1-\Omega_{\rm M}$.

\section{Observations, data reduction, and line extraction}
\label{sec:specobs}

\subsection{Single-dish IRAM-30m/EMIR spectroscopy}
\label{ssec:emirobs}

We observed high-density tracer molecules in the GEMS with EMIR on the IRAM 30-m telescope (project 108$-$14, PI Nesvadba). Observations
were carried out in June and September 2014, mostly with average weather conditions. We inspected all individual scans by eye, in particular,
those taken with the highest atmospheric opacity and precipitable water vapor (PWV) above 6\,mm. We flagged scans with spikes near the line
frequency or with unstable baselines. Most scans that were eventually discarded corresponded to high PWV values, but those taken at
${\rm PWV}>6\,$mm that had stable baselines were included.

We also searched for dense-gas emission lines falling in one of the sidebands during our previous observing runs with EMIR, which were
mainly designed to survey the CO and [\ion{C}{i}] emission lines in the E090 and E150 bands (see C18 and N19). We performed an automated
search using the best source redshifts and the local oscillator frequencies listed in the file headers. The resulting additional scans cover
numerous mid-$J$ lines of HCN, HCO$^+$, and HNC; these are listed in Table~\ref{tab:obslog} together with the corresponding project IDs.
We resampled scans from different observing runs onto a common frequency grid and, finally, we coadded them.

The tuning frequencies, integration times, and rms noise levels per spectral channel are listed in Table~\ref{tab:obslog}. We mainly used the
Wideband Line Multiple Autocorrelator (WILMA) backend, which offers more stable baselines,\footnote{In most cases, baseline breaks between the
  FTS units required us to apply baseline corrections and led to an increase in the spectrum rms.} apart from lines exclusively observed with
the Fast Fourier Transform Spectrometer (FTS; e.g., HNC(5--4) and HNC(7--6) in PLCK\_G092.5+42.9). We subtracted the baselines with first-order
polynomials. For non-detections, we computed the 3$\,\sigma$ upper limits on the flux, taking $\sigma$ as the uncertainty on the line intensity
defined in Eq.~1 of \citet{young11}. In these cases, WILMA provided systematically more constraining upper limits than FTS thanks to the
availability of better baselines. For each source, the fiducial full width at half maximum (FWHM) assumed for this calculation was the
unweighted averaged FWHM of all detected CO transitions \citepalias[see][]{canameras18b}. The number of line channels was obtained as
2\,$\times$\,FWHM/${v_{\rm res}}$, with ${v_{\rm res}}$ the channel width. We measured the spectrum rms in line-free regions, chosen using the
best-fitting redshift of the mid-$J$ CO lines, and after masking baseline channels not covered by all frequency setups. We list the flux upper
limits in Table~\ref{tab:linefit} for all transitions with a total integration time above two hours, except for those with more constraining
upper limits from NOEMA.

All lines detected at ${>}\,3\,\sigma$ were fitted with single Gaussian functions using the {\tt CLASS} software package from {\tt GILDAS}
\citep{gildas13}\footnote{http://www.iram.fr/IRAMFR/GILDAS}. The marginal detection of HCO$^+$(5--4) in PLCK\_G092.5+42.9 at signal-to-noise
ratio, ${\rm S/N \simeq 2.2}$, gives a velocity-integrated flux of $1.3 \pm 0.6\,{\rm Jy}\,{\rm km}\,{\rm s}^{-1}$, which is consistent with
NOEMA. Most detections in PLCK\_G244.8+54.9 have moderate S/N values of 3--5, and HNC(6--5) only reaches ${\rm S/N}=2.2$. These detections
are, nevertheless, considered significant because they result from careful baseline subtractions and channel masking, and because they are
systematically recovered when using subsets of the available spectral scans based on polarization or PWV. We also repeated the fit in spectral
regions that were expected to be line-free and did not detect any line at similar levels. Reduced EMIR spectra are shown in Fig.~\ref{fig:emir}.
Finally, we also stacked EMIR spectra covering emission lines of HCN, HCO$^+$, and HNC for the eight GEMS in Table~\ref{tab:obslog} to determine
whether or not this might provide more significant detections than for individual sources. The complete procedure is described in Appendix~A.
Only HCO$^+$(5--4) and HCO$^+$(6--5) have been robustly detected in the stacks and, consequently, we focus on line properties from individual
sources in the rest of the paper.

\begin{figure*}
  \includegraphics[width=0.37\textwidth]{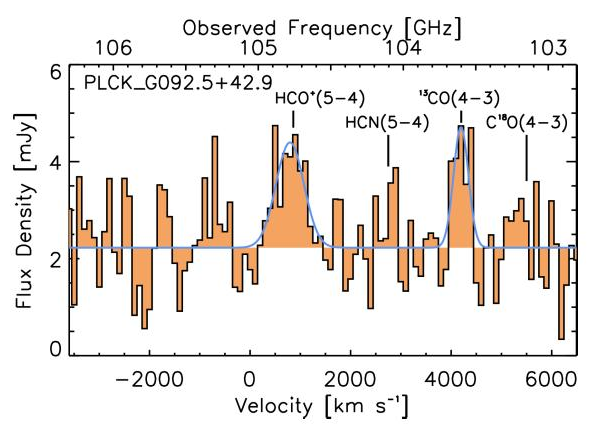}
  \includegraphics[width=0.32\textwidth]{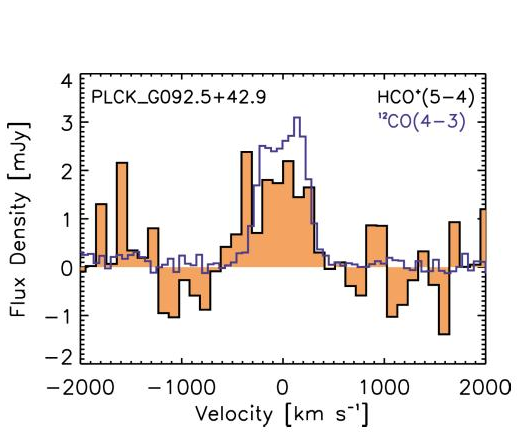}
  
  \includegraphics[width=0.37\textwidth]{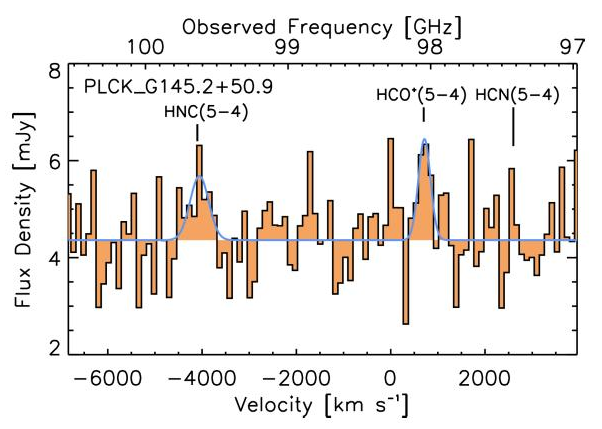}
  \includegraphics[width=0.32\textwidth]{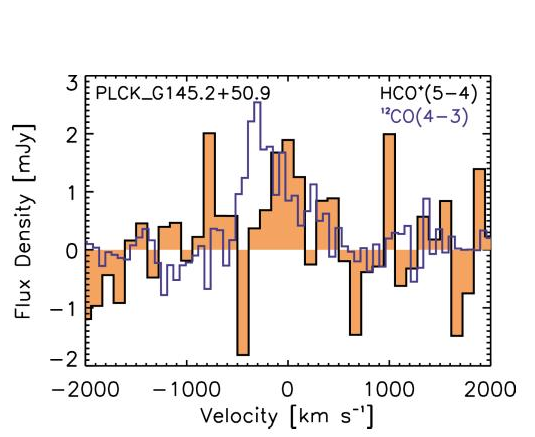}
  \includegraphics[width=0.32\textwidth]{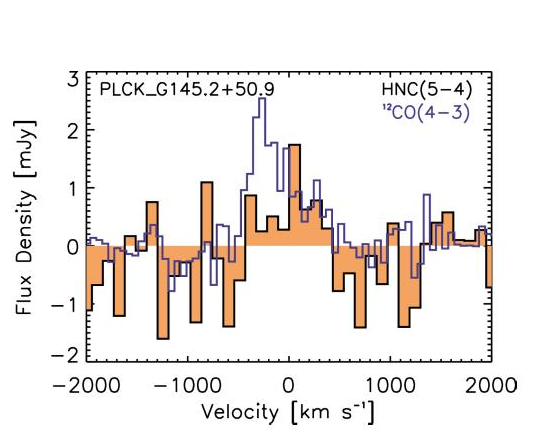}

  \includegraphics[width=0.37\textwidth]{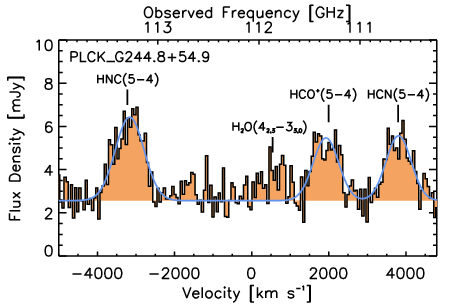}
  \includegraphics[width=0.32\textwidth]{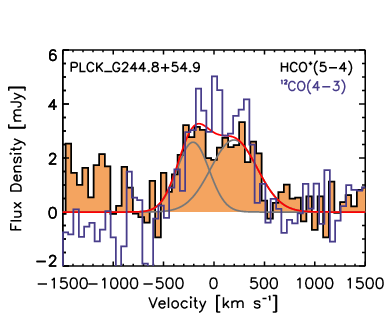}
  \includegraphics[width=0.32\textwidth]{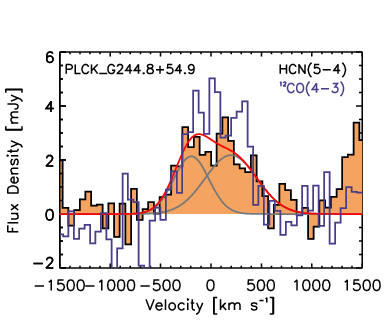}
  
  \caption{Spatially integrated NOEMA spectra of emission lines from dense-gas tracers binned to $100\,{\rm km}\,{\rm s}^{-1}$ resolution.
    {\it Left:} Best-fit Gaussian functions obtained with the {\tt mpfit} algorithm for all transitions detected at ${>}\,3\,\sigma$ (blue
    curves), together with the continuum. {\it Middle and right:} Profiles of individual lines compared with those of $^{12}$CO(4--3) (blue
    lines). CO spectra are rescaled arbitrarily and shifted according to the best spectroscopic redshifts from \citetalias{canameras18b}. For
    PLCK\_G244.8+54.9, the continuum-subtracted spectra of HCN(5--4) and HCO$^+$(5--4) were also fitted with two Gaussian components by fixing
    the peak velocities to those measured for CO lines \citepalias{canameras18b}. The best-fit line profiles are plotted as red curves and
    individual spectral components are overlaid as gray curves.}
  \label{fig:noema}
\end{figure*}

\begin{table*}
  \caption{Emission line properties and flux upper limits.}
\centering
\tiny
\begin{tabular}{lllllllll}
\hline
\hline
\noalign{\vskip 1pt}
Source & Line & Telescope & $\nu_{\rm obs}$ & Redshift & FWHM$_{\rm line}$ & $\mu I_{\rm line}$ & $\mu L_{\rm line}$ & $\mu L'_{\rm line}$ \\
\noalign{\vskip 1pt}
 & & & (GHz) & & (${\rm km}\,{\rm s}^{-1}$) & (${\rm Jy}\,{\rm km}\,{\rm s}^{-1}$) & (10$^{7}\,{\rm L}_{\odot}$) & (10$^{10}\,{\rm K}\,{\rm km}\,{\rm s}^{-1}\,{\rm pc}^2$) \\
\hline
\noalign{\vskip 1pt}
PLCK\_G092.5+42.9 & HCN(5--4) & NOEMA & \dots & \dots & (500) & $<$0.79 & $<$7.1 & $<$2.5 \\
 & HCO$^+$(5--4) & NOEMA & 104.75 $\pm$ 0.05 & 3.257 $\pm$ 0.007 & 742 $\pm$ 68 & 1.86 $\pm$ 0.16 & 16.8 $\pm$ 1.4 & 5.9 $\pm$ 0.5 \\
 & HNC(4--3) & IRAM/EMIR & \dots & \dots & (500) & $<$4.48 & $<$32.7 & $<$21.5 \\
 & HNC(5--4)$^{(*)}$ & \arcsec & \dots & \dots & (500) & $<$1.84 & $<$16.8 & $<$5.6 \\
 & HCN(7--6) & \arcsec & \dots & \dots & (500) & $<$1.52 & $<$19.0 & $<$2.5 \\
 & HCO$^+$(7--6) & \arcsec & \dots & \dots & (500) & $<$1.52 & $<$19.1 & $<$2.5 \\
 & HNC(7--6) & \arcsec & \dots & \dots & (500) & $<$1.81 & $<$23.1 & $<$2.8 \\
PLCK\_G145.2+50.9 & HCN(5--4) & NOEMA & \dots & \dots & (500) & $<$0.71 & $<$7.3 & $<$2.6 \\
 & HCO$^+$(5--4) & NOEMA & 98.02 $\pm$ 0.05 & 3.549 $\pm$ 0.008 & 271 $\pm$ 135 & 0.84 $\pm$ 0.20 & 8.7 $\pm$ 2.1 & 3.1 $\pm$ 0.7 \\
 & HNC(5--4)$^{(*)}$ & NOEMA & 99.60 $\pm$ 0.05 & 3.551 $\pm$ 0.008 & 429 $\pm$ 134 & 0.63 $\pm$ 0.19 & 6.6 $\pm$ 2.0 & 2.3 $\pm$ 0.7 \\
 & HCN(7--6) & IRAM/EMIR & \dots & \dots & (500) & $<$1.57 & $<$22.6 & $<$3.0 \\
 & HCO$^+$(7--6) & \arcsec & \dots & \dots & (500) & $<$1.57 & $<$22.7 & $<$2.9 \\
 & HNC(7--6) & \arcsec & \dots & \dots & (500) & $<$1.59 & $<$23.4 & $<$2.9 \\
PLCK\_G244.8+54.9 & HCN(5--4) & NOEMA & 110.61 $\pm$ 0.01 & 3.006 $\pm$ 0.001 & 759 $\pm$ 47 & 2.40 $\pm$ 0.20 & 18.9 $\pm$ 1.6 & 6.9 $\pm$ 0.6 \\
 & HCO$^+$(5--4) & NOEMA & 111.31 $\pm$ 0.01 & 3.006 $\pm$ 0.001 & 768 $\pm$ 33 & 2.41 $\pm$ 0.16 & 19.0 $\pm$ 1.3 & 6.9 $\pm$ 0.5 \\
 & HNC(5--4)$^{(\dag)}$ & ALMA & (113.16) & (3.005) & 535 $\pm$ 101 & 2.24 $\pm$ 0.49 & 18.0 $\pm$ 3.9 & 6.0 $\pm$ 1.3 \\
 & CN(4--3)$^{(\dag)}$ & ALMA & (113.25) & (3.005) & 668 $\pm$ 70 & 1.27 $\pm$ 0.34 & 10.1 $\pm$ 2.7 & 3.4 $\pm$ 0.9 \\
 & HCN(4--3) & IRAM/EMIR & 88.52 $\pm$ 0.03 & 3.005 $\pm$ 0.004 & 631 $\pm$ 157 & 1.73 $\pm$ 0.47 & 10.8 $\pm$ 2.9 & 7.6 $\pm$ 2.1 \\
 & HCO$^+$(4--3) & \arcsec & 88.99 $\pm$ 0.04 & 3.009 $\pm$ 0.005 & 822 $\pm$ 376 & 2.25 $\pm$ 0.70 & 14.2 $\pm$ 4.4 & 9.8 $\pm$ 3.0 \\
 & HNC(4--3) & \arcsec & 90.51 $\pm$ 0.04 & 3.007 $\pm$ 0.005 & 769 $\pm$ 252 & 1.63 $\pm$ 0.53 & 10.4 $\pm$ 3.4 & 6.9 $\pm$ 2.2 \\
 & HCN(6--5) & \arcsec & \dots & \dots & (650) & $<$1.80 & $<$17.2 & $<$3.1 \\
 & HCO$^+$(6--5) & \arcsec & \dots & \dots & (650) & $<$1.80 & $<$17.2 & $<$3.1 \\
 & HNC(6--5) & \arcsec & (135.79) & (3.005) & (650) & 1.34 $\pm$ 0.61 & 12.8 $\pm$ 5.8 & 2.4 $\pm$ 1.1 \\[+0.5em]
PLCK\_G045.1+61.1 & HCN(5--4) & IRAM/EMIR & \dots & \dots & (650) & $<$4.54 & $<$44.0 & $<$15.8 \\
 & HCO$^+$(5--4) & \arcsec & \dots & \dots & (650) & $<$4.54 & $<$44.3 & $<$15.6 \\
PLCK\_G113.7+61.0 & HCN(4--3) & \arcsec & \dots & \dots & (500) & $<$4.19 & $<$18.2 & $<$12.8 \\
 & HCO$^+$(4--3) & \arcsec & \dots & \dots & (500) & $<$4.74 & $<$20.7 & $<$14.3 \\
 & HNC(4--3) & \arcsec & \dots & \dots & (500) & $<$4.74 & $<$21.0 & $<$13.8 \\
PLCK\_G138.6+62.0 & HCN(4--3) & \arcsec & \dots & \dots & (500) & $<$5.35 & $<$23.6 & $<$16.6 \\
 & HCO$^+$(4--3) & \arcsec & \dots & \dots & (500) & $<$5.35 & $<$23.8 & $<$16.4 \\
PLCK\_G165.7+67.0 & HCN(6--5) & \arcsec & \dots & \dots & (600) & $<$3.02 & $<$17.2 & $<$3.6 \\
 & HCO$^+$(6--5) & \arcsec & \dots & \dots & (600) & $<$3.02 & $<$17.3 & $<$3.5 \\
 & HNC(6--5) & \arcsec & \dots & \dots & (600) & $<$2.27 & $<$13.2 & $<$2.6 \\
PLCK\_G200.6+46.1 & HCN(4--3) & \arcsec & \dots & \dots & (450) & $<$2.60 & $<$16.0 & $<$11.2 \\
 & HCO$^+$(4--3) & \arcsec & \dots & \dots & (450) & $<$2.60 & $<$16.1 & $<$11.1 \\
\hline
\end{tabular}
\tablefoot{The emission line fluxes and luminosities were obtained by fitting the spectra with single Gaussians, either with {\tt CLASS} or
  our own fitting routines. We list the 3$\,\sigma$ flux upper limits for all transitions observed in one of the sidebands for more than two
  hours. Columns are: source name; transition; telescope; observed frequency; redshift of the line or spectral component; line FWHM;
  observed velocity-integrated flux density $\mu I_{\rm line}$ used for the line excitation analysis; and observed line luminosity in solar
  luminosities and in ${\rm K}\,{\rm km}\,{\rm s}^{-1}\,{\rm pc}^2$ units. The values of line fluxes and luminosities are not corrected
  for lensing magnification ($\mu I_{\rm line}$ refers to the intrinsic source-plane flux times the magnification factor). Asterisks
  indicate that HNC(5--4) is blended with CN(4--3), with a velocity offset of about 300\,km\,s$^{-1}$, and that listed fluxes and
  upper limits do not distinguish these two components. Daggers indicate values corrected for 30\% missing flux due to the small
  0.14\arcsec\,$\times$\,0.06\arcsec\ beam. The FWHM values in brackets are the averaged FWHM of all detected CO transitions from
    \citetalias{canameras18b}.}
\label{tab:linefit}
\end{table*}

\subsection{Spectral-line interferometry}
\label{ssec:interfobs}

\subsubsection{NOEMA}

The two GEMS, PLCK\_G092.5+42.9 and PLCK\_G145.2+50.9, were observed with NOEMA in September and October 2015 (project S15CH, PI Ca\~nameras).
For each track, six antennas were included in the D configuration observations, with baselines between 24\,m and 97\,m. The phase centers
of the observations were located at $\alpha$=16:09:17.76 and $\delta$=+60:45:21.0 for PLCK\_G092.5+42.9, and at $\alpha$=10:53:22.56 and
$\delta$=+60:51:49.0 for PLCK\_G145.2+50.9. The 3-mm receivers were tuned to cover the sky frequencies of the HCN and HCO$^+$\,$J=5$--4
lines for both GEMS. The WideX correlator was used as the backend, with its 3.6-GHz bandwidth at a spectral resolution of 1.95\,MHz.

For PLCK\_G092.5+42.9, \object{MWC~349} and \object{LkHa~101} were used for bandpass calibration, while the bright quasars \object{3C454.3},
\object{0851+202}, \object{1749+096}, and \object{2013+370} were observed as primary flux calibrators, and we regularly observed the nearby
quasars \object{1642+690} and \object{1637+574} for gain and phase calibration. For PLCK\_G145.2+50.9, \object{LkHa~101} and \object{3C84}
were used for bandpass and flux calibration, while \object{1030+611} and \object{1044+719} were used for gain and phase calibration.

The data were calibrated and imaged within {\tt GILDAS/CLIC} and {\tt MAPPING}. To image the data, we applied natural weighting,
resulting in beamsizes of 6.1\arcsec\,$\times$\,4.0\arcsec\ at a position angle of 67.6\degr\ for PLCK\_G092.5+42.9, and
5.6\arcsec\,$\times$\,4.3\arcsec\ at a position angle of 118.9\degr\ for PLCK\_G145.2+50.9. We produced data cubes of 35-MHz spectral
resolution, with channel spacings of $99.9\,{\rm km}\,{\rm s}^{-1}$ and $106.8\,{\rm km}\,{\rm s}^{-1}$ for PLCK\_G092.5+42.9 and
PLCK\_G145.2+50.9, respectively.

In addition, $J=5$--4 lines of HCN, HCO$^+$, and HNC for PLCK\_G244.8+54.9 were observed in December 2019, with a beamsize of
2.9\arcsec\,$\times$\,1.2\arcsec\ at a position angle of 14.2\degr, as part of a broader spectral window (project S19CU, PI Nesvadba).
The corresponding line fluxes were added to our current analysis, while the overall data will be presented in a forthcoming paper.

We derived spatially integrated spectra by coadding spatial pixels above 3$\,\sigma$ in velocity-integrated maps to maximize the S/N
(see Fig.~\ref{fig:noema}). All lines detected at more than 3$\,\sigma$ were fitted simultaneously with Gaussian profiles using the {\tt mpfit}
algorithm, together with the continuum. Their properties are listed in Table~\ref{tab:linefit}. We computed uncertainties using Monte Carlo
simulations, by randomly drawing the flux in each spectral channel assuming flux uncertainties of 10\% and reproducing the fitting procedure.
Errors on the line parameters correspond to the mean and standard deviation of the distributions obtained after 1000 iterations.

Defining the signal-to-noise ratios as $I_{\rm peak} / \Delta I_{\rm peak}$, with $I_{\rm peak}$ the best-fitting peak flux density and
$\Delta I_{\rm peak}$ its uncertainty, the detections of HCO$^+$(5--4) and $^{13}$CO(4--3) in PLCK\_G092.5+42.9 have S/N values of 9.5$\,\sigma$
and 9.0$\,\sigma$, respectively. In PLCK\_G145.2+50.9, HNC(5--4) is detected at 3.8$\,\sigma$ and HCO$^+$(5--4) at 3.5$\,\sigma$. These values
are consistent with line fluxes measured on low S/N EMIR detections. Finally, in PLCK\_G244.8+54.9, HCN(5--4), HCO$^+$(5--4), and HNC(5--4) have
S/N values of 14.3$\,\sigma$, 15.5$\,\sigma$, and 19.5$\,\sigma$, respectively. We also measure a 3-mm dust continuum of $(2.26\pm0.04)$\,mJy
at $\simeq 104$\,GHz in PLCK\_G092.5+42.9, $(4.42\pm0.09)\,$mJy at $\simeq 98$\,GHz in PLCK\_G145.2+50.9, and $(2.51\pm0.05)\,$mJy
at $\simeq 112$\,GHz in PLCK\_G244.8+54.9.

\subsubsection{ALMA}

For PLCK\_G244.8+54.9, we also used ALMA data obtained in Cycle 3 (2015.1.01518.S, PI Nesvadba). These correspond to two science goals in
Band~3, with lower and higher spatial resolution, respectively. 

The lower-resolution data were taken on 21 July 2016, with 39 antennas (configuration C40-5) and specifically targeted dense-gas line tracers
in three basebands, that is, HNC(5--4) centered at a redshifted frequency of 113.115\,GHz, HCN(5--4) at 110.635\,GHz, and $^{13}$CO(4--3) at
110.046\,GHz. We used narrower basebands, of $\lesssim$1200~km~s$^{-1}$, than we had for the higher-resolution data. For the reduction, we used
  standard manual scripts with the Common Astronomy Software Application \citep[{\tt CASA},][]{mcmullin07}. The bandpass and amplitude/flux
calibrator was J1058+0133 and the phase calibrator was J1041+0610, 3.1\degr\ from PLCK\_G244.8+54.9. Continuum subtraction was complicated by
the fact that the ALMA bandwidth is almost completely covered by line emissions. We identified a part of the fourth ALMA baseband centered on
111.616\,GHz as the most weakly contaminated range and used it as a spatial template with lowest line emission. However, we needed to
correct the amplitude of the continuum template map by a factor 0.67 to account for the integrated line flux density of
H$_{\rm 2}$O(4$_{\rm 2,3}$--3$_{\rm 3,0}$) at 111.84\,GHz based on the NOEMA spectrum of PLCK\_G244.8+54.9 in Fig.~\ref{fig:noema}. We produced
the data cubes for the final analysis using {\tt CLEAN} ``Briggs'' weighting with ${\rm robust=1.0}$ and the synthesized beamsize is
$0.78\arcsec\times0.67\arcsec$ at a position angle of $-83.6$\degr.

We used these lower angular resolution, continuum-subtracted line cubes to derive spatially-integrated spectra and line maps of HCN(5--4)
and HNC(5--4), with a channel width of 15\,MHz. We coadded frequency channels covering these emission lines to produce velocity-integrated
maps and summed up pixels above 3$\,\sigma$ to obtain integrated spectra. Emission lines were fitted with single Gaussians and errors computed
with flux uncertainties per spectral channel of 10\%. The HCN(5--4) line flux of $(2.05\pm0.25)\,{\rm Jy}\,{\rm km}\,{\rm s}^{-1}$ is consistent
within 1$\,\sigma$ with that from NOEMA, and the HNC(5--4) flux matches those from the higher-resolution data cube and from NOEMA within
1$\,\sigma$. Line profiles are also comparable. Consequently, given that these uncertainties and the limited width of the HCN(5--4) and HNC(5--4)
basebands (${<}\,2\times{\rm FWHM}$) are insufficient for producing a robust characterization of the baseline levels and line profiles, we focus
the following discussion on the highest S/N detections of HCN(5--4) and HNC(5--4) from NOEMA and from the higher-resolution ALMA data cube,
respectively.

These higher resolution observations targeted the redshifted CO(4--3) line centered at 114.888\,GHz (i.e., $z=3.005$) with high spectral
resolution (see \citetalias{canameras17b}); it also covered the HNC(5--4) line in a baseband centered at 112.897\,GHz, with a channel width
of 15.6\,MHz. These data were taken on 23 October 2015, with 35 antennae (configuration C36-8) in good conditions with PWV of around
0.6--1.0\,mm and high phase stability (95--100\,$\mu$m rms on baselines of 6500\,m). We applied automatic and manual flagging of visibilities,
calibrating bandpass and amplitude/flux with J1058+0133, and phase with J1044+0655, 2.4\degr\ from PLCK\_G244.8+54.9. We used a ``Briggs''
weighting with ${\rm robust=1.0}$ and obtained a synthesized beam of $0.119\arcsec\times0.081\arcsec$ at a position angle of 45.3\degr. Another
source, J1038+0512, which is 2.3\degr\ from J1044+0655, was also included in the scan sequence to allow us to verify the quality of the transfer
of the phase solutions from the phase calibrator. The rms in the data cube is $0.18\,{\rm mJy}\,{\rm beam}^{-1}$, with a spectral channel width
of $42\,{\rm km}\,{\rm s}^{-1}$.

\begin{figure}
  \centering
    \includegraphics[width=0.24\textwidth]{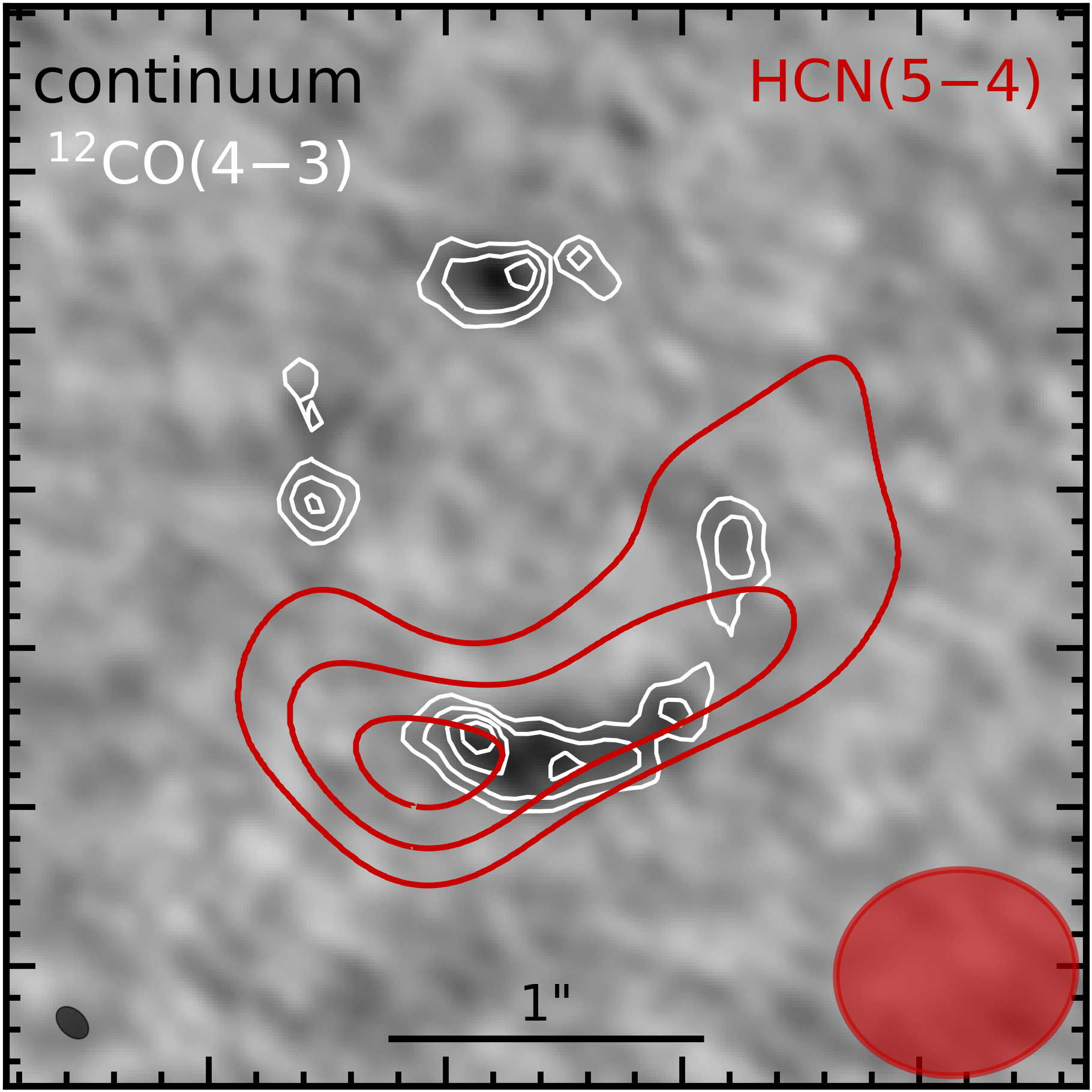}
    \includegraphics[width=0.24\textwidth]{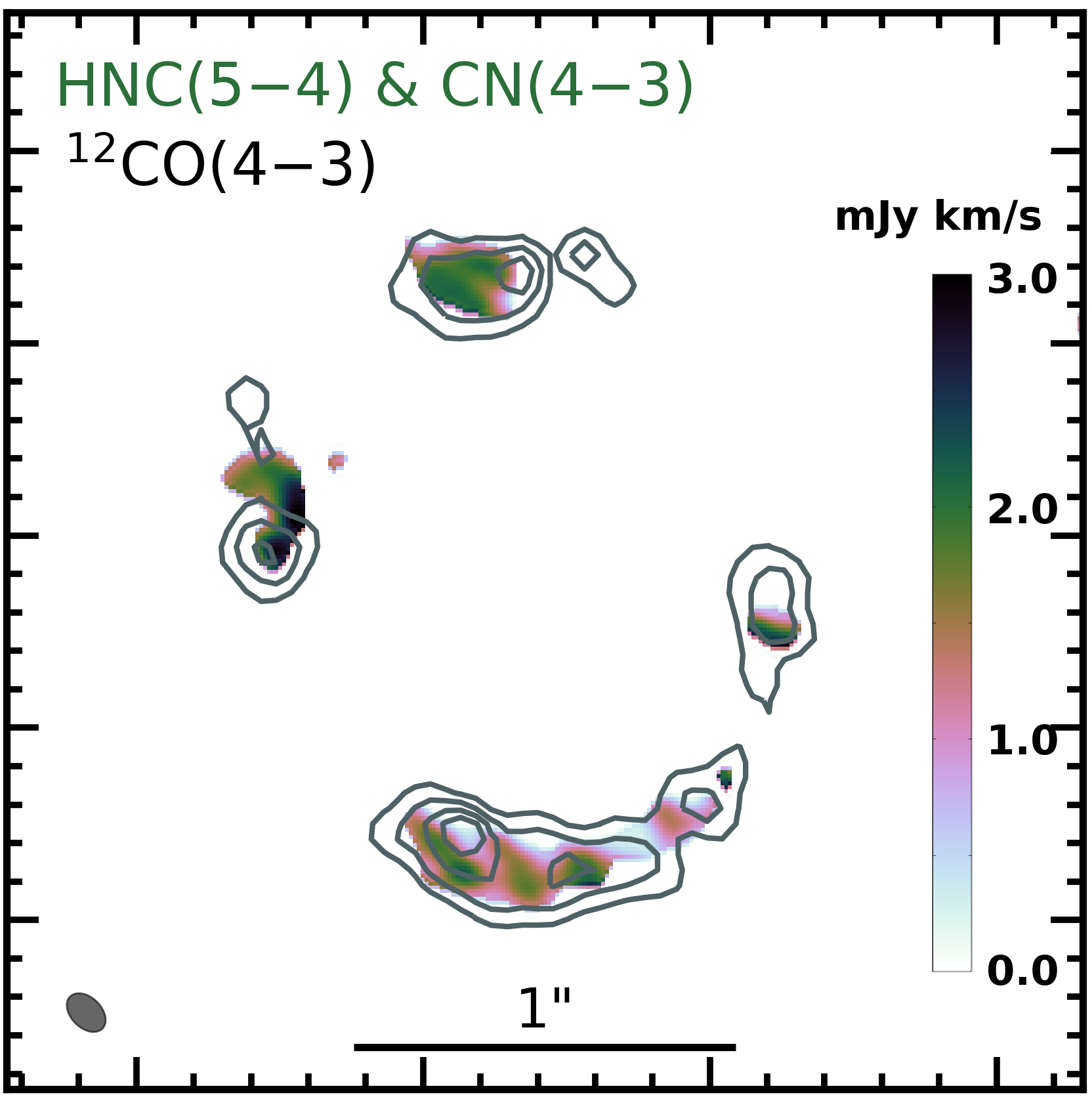}
    \caption{{\it Left:} Velocity-integrated HCN(5--4) map of PLCK\_G244.8+54.9 from ALMA (red contours) with a beamsize of
        $0.78\arcsec\times0.67\arcsec$ at ${\rm PA=-83.6}$\degr (red ellipse in the lower-right corner), compared with the CO(4--3) line
        emission (white contours) and 3-mm dust continuum (grey background) at 0.1\arcsec\ resolution from \citetalias{canameras17b}. Contour
        levels are 4$\sigma$, 6$\sigma$, 8$\sigma$ (red) and 3$\sigma$, 6$\sigma$, 9$\sigma$ (white). {\it Right:} Zoom onto the morphology of
        the velocity-integrated HNC(5--4) line emission blended with CN(4--3) (see Sect.~\ref{ssec:deblending}). The black contours correspond
        to the white ones in the left panel. Local velocity-integrated flux densities of the blended HNC(5--4) line are shown as a color map,
        with the same beamsize of $0.119\arcsec\times0.081\arcsec$ as CO(4--3) (grey ellipse in the lower-left corner).}
  \label{fig:almamaps}
\end{figure}

We created continuum and HNC(5--4) line-flux maps from the higher-resolution data cube by fitting the line profile in spectra extracted over
circular apertures of 0.08\arcsec\ diameter, which is slightly lower than the beamsize, in order to maximize the S/N without losing spatial
information. For each pixel with ${\rm S/N}\,{>}\,3$ on the velocity-integrated line map, we fitted a single Gaussian component using
{\tt mpfit}. Given the relatively low S/N per aperture, we fitted only the peak flux by keeping the line centroid and FWHM fixed to values
measured on the integrated spectrum (see Table~\ref{tab:linefit}), and the continuum fixed to the average over the line-free channels. The
resulting maps are shown in Fig.~\ref{fig:almamaps}.

The HNC(5--4) line profile is characterized using a ($u$,$v$) tapered version of the higher-resolution data cube, with beamsize of
$0.21\arcsec\times0.16\arcsec$ at ${\rm PA=53.7^{\circ}}$. This cube improves the spectrum S/N, spatially resolves the partial Einstein ring, and
has an rms of $0.24\,{\rm mJy}\,{\rm beam}^{-1}$ per spectral channel. We created a velocity-integrated line map and summed up pixels above
3$\,\sigma$ from this map. The resulting source-integrated spectrum is shown in Fig.~\ref{fig:alma} and has a rms level of 0.5\,mJy. Similarly
to \citetalias{canameras17b}, we recover about 70\% of the source-integrated continuum flux density at 3-mm expected from our modified blackbody
fit of the FIR/sub-mm spectral energy distribution in \citetalias{canameras15}. The missing emission is likely from extended regions filtered
out with this higher-resolution configuration and has been added to the observed HNC(5--4) line flux in Table~\ref{tab:linefit}.

\section{Emission line properties}
\label{sec:lineprop}

In Table~\ref{tab:linefit}, we list properties of all the emission lines from high-density tracer molecules observed with EMIR, NOEMA, and ALMA,
including the line FWHMs, velocity-integrated fluxes, and luminosities obtained with single Gaussian fits (or the 3$\,\sigma$ upper limits).
In the subsequent analysis and discussion, we only consider line properties and line ratios in PLCK\_G092.5+42.9, PLCK\_G145.2+50.9, and
PLCK\_G244.8+54.9 (which are the GEMS with the deepest integrations), while the others are not detected individually in dense-gas tracers. For
these three sources, ${\rm rms}\,{<}\,0.4\,$mK on EMIR spectra provide particularly tight flux upper limits.

\subsection{Line profiles}

\begin{figure}
  \centering
    \includegraphics[width=0.40\textwidth]{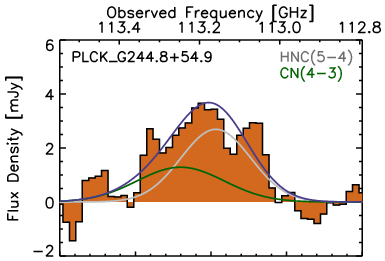}  

    \includegraphics[width=0.40\textwidth]{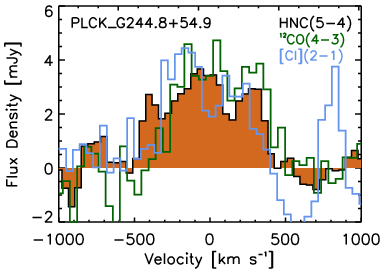}
    \caption{Spatially integrated ALMA spectrum of HNC(5--4) in PLCK\_G244.8+54.9, with a spectral channel width of 42\,km\,s$^{-1}$.
      {\it Top:} Dark blue line shows the best-fit profile, with the deblended HNC(5--4) and CN(4--3) emission lines plotted in grey
      and green, respectively (see Sect.~\ref{ssec:deblending}). {\it Bottom:} Comparison of the line profile with those of CO(4--3)
      and [\ion{C}{i}](2--1).}
  \label{fig:alma}
\end{figure}

Figures~\ref{fig:emir}, \ref{fig:noema}, and \ref{fig:alma} show the profiles of emission lines detected at more than 3$\,\sigma$ with
the IRAM-30m/EMIR, NOEMA, and ALMA. All spectra except those showing the complete NOEMA spectral windows have been continuum subtracted
using first-order polynomials fitted to line-free spectral channels. Due to their limited S/N, the EMIR spectra of dense-gas tracers cannot
distinguish individual velocity components. Those for which we fitted the HCN, HCO$^+$, and HNC line FWHMs suggest comparable line
widths with those of $^{12}$CO (CO hereafter) in \citetalias{canameras18b}, despite large uncertainties. Line widths in PLCK\_G244.8+54.9
are large but remain consistent with the range of 580--$750\,{\rm km}\,{\rm s}^{-1}$ of high S/N lines across the CO ladder
($J_{\rm up}=3$--11)\footnote{For single Gaussian fits not listed in C18.}, suggesting that these tracers probe highly turbulent gas reservoirs
with similar spatial distributions.

Emission lines detected with NOEMA and ALMA have higher S/N and better resolution than those observed with EMIR, and most of them are well
fitted by single Gaussians. They are overplotted in Figs.~\ref{fig:noema} and \ref{fig:alma} with the mid-$J$ CO transition from
\citetalias{canameras18b} with the highest S/N, shifted to the same frequency, rescaled arbitrarily, and rebinned to spectral resolutions of
100 and $40\,{\rm km}\,{\rm s}^{-1}$, respectively (rightmost panels of Fig.~\ref{fig:noema} and bottom panel of Fig.~\ref{fig:alma}). The
widths and centroids of dense-gas and CO transitions are similar in most cases, in line with previous findings for high-redshift SMGs
\citep{oteo17,bethermin18}. For instance, the profiles of HCO$^+$(5--4) and CO(4--3) in PLCK\_G092.5+42.9 match reasonably well. For
PLCK\_G145.2+50.8, the HCO$^+$(5--4) profile seems asymmetric and shifted by about $200\,{\rm km}\,{\rm s}^{-1}$ with respect to CO(4--3)
\cite[akin to HCN(3--2) in the Cosmic Eyelash,][]{danielson11}, with a low FWHM of $(271\pm135)\,{\rm km}\,{\rm s}^{-1}$. These variations
might be intrinsic, suggesting more compact dense-gas emitting regions than the bulk of the CO gas reservoirs, but we caution that the
deficit of HCO$^+$(5--4) emission near the CO(4--3) peak needs to be confirmed with a higher S/N spectrum.

The HCN(5--4) and HCO$^+$(5--4) emission lines in PLCK\_G244.8+54.9 also have similar widths and centroids to those of mid- and high-$J$
CO, for single Gaussian fits; Fig.~\ref{fig:noema} illustrates how well-matched these profiles are. It is worth noting that these
spatially-integrated spectra also closely resemble that of HNC(5--4) from ALMA in Fig.~\ref{fig:alma}, when excluding the blue emission
tail from $-400$ to $-250\,{\rm km}\,{\rm s}^{-1}$ (likely due to contamination of HNC(5--4) by CN(4--3), as demonstrated in
Sect.~\ref{ssec:deblending}). HCN, HCO$^+$, and HNC should therefore probe nearly co-spatial regions in PLCK\_G244.8+54.9, which is consistent
with recent studies on the inner regions of local star-forming galaxies \citep[e.g., HCN and HCO$^+$ $J=4$--3 in the central few kpc of NGC\,253,
  NGC\,1068, IC\,342, M\,82, M\,83, and NGC\,6946,][]{tan18}. Moreover, the similarity between dense-gas and CO lines in PLCK\_G244.8+54.9
suggests that the strongly magnified region of $\sim$1~kpc source-plane size \citepalias{canameras17b} is fairly homogeneous in line
properties, in spite of its outstanding star formation and gas-mass surface densities. This region likely hosts enough dense gas to sample
the same velocity profile as seen in CO. This argues for a genuine multiphase medium, with ongoing mass exchange between dense and more
diffuse gas driven by turbulence.

Overall, we obtained the lowest dense-gas line fluxes and the most significant differences between line profiles in PLCK\_G145.2+50.8. This
could mean that gas phases are better mixed in PLCK\_G244.8+54.9 (and perhaps in PLCK\_G092.5+42.9) than in PLCK\_G145.2+50.8, where our data
may cover fewer dense cores and more diffuse gas.

\subsection{Individual spectral components}

For the three GEMS in the current analysis, CO lines covered in C18 exhibit double-peaked profiles with individual spectral components 
offset by 300 to 450~km~s$^{-1}$ and with $<$50\% variations in their flux ratios \citepalias[see description in][]{canameras18b}. We cannot
directly measure the dense-gas emission associated with each of these components for PLCK\_G092.5+42.9 and PLCK\_G145.2+50.9 using EMIR and
NOEMA. However, for PLCK\_G092.5+42.9, Fig.~\ref{fig:noema} qualitatively suggests that intensity ratios between dense tracers and CO
are comparable in the red and blue spectral features separated by $350\,{\rm km}\,{\rm s}^{-1}$. For PLCK\_G145.2+50.9, NOEMA spectra point
towards lower HCO$^+$(5--4)/CO(4--3) and HNC(5--4)/CO(4--3) in the blue kinematic component, but these have low S/N.

For PLCK\_G244.8+54.9, the NOEMA spectra of HCN(5--4) and HCO$^+$(5--4) are reasonably well fitted with single Gaussians but show a common
excess around $-300\,{\rm km}\,{\rm s}^{-1}$ (Fig.~\ref{fig:noema}). This suggests double-peaked profiles with line components separated by
$\lesssim 400\,{\rm km}\,{\rm s}^{-1}$, akin to all spatially integrated CO spectra with $J_{\rm up}=3$--10 \citepalias{canameras18b} for this
source and to the CO(4--3) line profiles in multiple images across the source from our high-resolution study \citepalias{canameras17b}. We
fitted these two kinematic components by fixing their central velocity to the average CO line centroids over the SLED of $-170$ and
$+190\,{\rm km}\,{\rm s}^{-1}$. The blue and red spectral components of HCO$^+$(5--4) have ${\rm FWHM} = (283\pm36)\,{\rm km}\,{\rm s}^{-1}$
and $I = (0.84 \pm 0.13)\,$mJy, and ${\rm FWHM} = (449\pm91)\,{\rm km}\,{\rm s}^{-1}$ and $I = (1.32 \pm 0.16)\,$mJy, respectively. The red
component is broader by a factor of approximately 1.6, which is in excellent agreement with FWHM ratios in the range 1.4--1.9 measured on
single-dish spectra of mid- and high-$J$ CO lines \citepalias{canameras18b}; the red component is also brighter. The blue spectral component
of HCN(5--4) has ${\rm FWHM} = (385\pm85)\,{\rm km}\,{\rm s}^{-1}$ and $I = (0.98 \pm 0.18)\,$mJy, while the red component is 1.3 times broader,
with ${\rm FWHM} = (486\pm107)\,{\rm km}\,{\rm s}^{-1}$ and $I = (1.27 \pm 0.28)\,$mJy. We did not attempt to fit the components separately for
HNC(5--4), which is blended with CN(4--3). The similar FWHM and flux ratios found between these individual Gaussians indicate that mid-$J$
CO lines, HCN, and HCO$^+$ do not trace gas reservoirs with fundamentally different kinematics in PLCK\_G244.8+54.9. This also suggests that
line-emitting regions are nearly collocated in the source plane within the rotating disk probed in CO(4--3) \citepalias{canameras17b}.

\begin{table*}
  \caption{Flux ratios of dense-gas tracers to CO and [\ion{C}{i}] lines.}
\centering
\begin{tabular}{llllllll}
\hline
\hline
\rule{0pt}{3ex} Source & $I_{\rm HCN}/I_{\rm CO\ mid-J}$ & $I_{\rm HCO^+}/I_{\rm CO\ mid-J}$ & $I_{\rm HNC}/I_{\rm CO\ mid-J}$ & $I_{\rm HCN}/I_{\rm CO10}$ & $I_{\rm HCO^+}/I_{\rm CO10}$ & $I_{\rm HCN}/I_{\rm [\ion{C}{i}]10}$ & $I_{\rm HCO^+}/I_{\rm [\ion{C}{i}]10}$ \\[+0.5em]
\hline
\noalign{\vskip 1pt}
PLCK\_G092.5+42.9 & $<$0.022 & 0.052 $\pm$ 0.007 & $<$0.031 & $<$0.10 & 0.24 $\pm$ 0.09 & $<$0.059 & 0.14 $\pm$ 0.04 \\
PLCK\_G145.2+50.9 & $<$0.015 & 0.018 & 0.008 & $<$0.11 & 0.12 & $<$0.05 & 0.06 $\pm$ 0.03 \\
PLCK\_G244.8+54.9 & 0.064 $\pm$ 0.023$\dag$ & 0.084 $\pm$ 0.034$\dag$ & 0.061 $\pm$ 0.025$\dag$ & 0.40 $\pm$ 0.23$\dag$ & 0.52 $\pm$ 0.30$\dag$ & 0.07$\dag$ & 0.09$\dag$ \\
 & 0.071 $\pm$ 0.009 & 0.072 $\pm$ 0.008 & 0.066 $\pm$ 0.018 & 0.56 $\pm$ 0.22 & 0.56 $\pm$ 0.21 & 0.10 & 0.10 \\
\hline
\end{tabular}
\tablefoot{Dense-gas to CO and [\ion{C}{i}] flux ratios derived from values in Table~\ref{tab:linefit}, from the line measurements of
  \citetalias{canameras18b} and \citetalias{nesvadba19}, and from \citet{harrington18}. All rotational levels are $J=5$--4, except for
  PLCK\_G244.8+54.9 where ratios marked with a dagger correspond to $J=4$--3. We do not list uncertainties when the [\ion{C}{i}] or CO
  line flux is converted from another transition (see Sect.~\ref{ssec:lineratios}). CO(1--0) in PLCK\_G145.2+50.9 is extrapolated
  using the average ratio $I_{\rm CO(4-3)}/I_{\rm CO(1-0)} \simeq 5.1$ in the GEMS.}
\label{tab:lineratios}
\end{table*}

\subsection{Deblending HNC(5--4) and CN(4--3)}
\label{ssec:deblending}

The HNC(5--4) emission line is blended with the $N=4$--3 transition of the cyanide radical CN (CN(4--3) hereafter), with a velocity offset of
about $300\,{\rm km}\,{\rm s}^{-1}$. Here, CN(4--3) has multiple components due to fine and hyperfine splitting, and its two most prominent
components are within $150\,{\rm km}\,{\rm s}^{-1}$ and have comparable intensities \citep{guelin07}. At high redshift, HNC(5--4) and CN(4--3)
have been detected in the stacked spectrum of SPT lensed SMGs \citep[][]{spilker14}, albeit with spectral resolution that was too limited to
directly measure the individual line fluxes. While HNC(5--4) is expected to be brighter (see also below), significant contamination was
previously measured in a few high-redshift sources, including the lensed quasar APM~08279+5255 \citep{guelin07} and two high-$z$ lensed SMGs
\citep{bethermin18}. This demonstrates the need to perform a careful deblending of these lines.

For PLCK\_G244.8+54.9, the HNC(5--4) line shown in Fig.~\ref{fig:alma} is broad and exhibits an excess of blueshifted emission around
$-400\,{\rm km}\,{\rm s}^{-1}$, which is not observed in the [\ion{C}{i}](2--1) and CO(4--3) line profiles and does not match the expected
velocities of the independent red and blue spectral components (offset by around $400\,{\rm km}\,{\rm s}^{-1}$ from each other, see C18).
The flux excess rather corresponds to the redshifted frequency of CN(4--3), similarly to \citet{guelin07}, suggesting that CN(4--3) makes
a non-negligible contribution to the combined line flux.

We deblended the two emission lines using our highest S/N ALMA spectrum (Fig.~\ref{fig:alma}) by fitting them simultaneously, with peak
frequencies fixed to our best redshift estimate from CO \citepalias{canameras18b}.\footnote{We obtained consistent results using the
  HNC(5--4)+CN(4--3) detection from NOEMA with lower S/N.} To derive robust constraints on the line FWHMs and fluxes, we fitted each
transition with a single Gaussian without considering neither independent velocity components in this source, nor the individual
CN(4--3) components. The resulting line FWHMs are similar to those of $J=5$--4 and $J=4$--3 transitions of high-density tracers
in PLCK\_G244.8+54.9 (see Table~\ref{tab:linefit}), and we obtained flux uncertainties of about 20\%. The HNC(5--4)/CN(4--3) integrated
flux ratio is 1.76$\pm$0.86, remarkably similar to measurements from the literature, including the APM~08279+5255 quasar \citep[around
  1.7,][]{guelin07}, and individual strongly lensed SPT SMGs at $z\simeq2.5$--3.0 \citep[around 1.6,][]{bethermin18}. Given that
for PLCK\_G145.2+50.9, the spectrum S/N is too low to estimate individual line fluxes, we used the same ratio to infer an HNC(5--4) flux
of 0.40\,Jy\,km\,s$^{-1}$ from the blended line flux given in Table~\ref{tab:linefit}.

\subsection{Discussion on differential lensing}

High-density tracer molecules are expected to reside in the inner core of giant, star-forming molecular clouds and could potentially probe
an ISM component that is distinct from that probed in the low-to-mid-$J$ CO, or [\ion{C}{i}] lines. Different spatial distributions
have been observed in low-redshift (U)LIRGs \citep[see, e.g.,][]{scoville15,scoville17,konig18}, where dense regions are located within the
starburst cores and embedded in diffuse, colder gas reservoirs. High-redshift lensed SMGs could also have such configurations and, in this
case, spatially integrated emission line fluxes and profiles could be strongly affected by differential magnification
\citep{hezaveh12,serjeant12}, introducing high uncertainties on the diagnostics from line ratios.

We used the high-spatial resolution HNC(5--4) cube from ALMA for PLCK\_G244.8+54.9 to compare the intensity-weighted magnification factor
of this line with that of CO(4--3) \citepalias[from][]{canameras17b}. These exquisite data allow us to directly study the spatial
distribution of molecular gas phases probed by the two transitions, for the first time at high redshift, thanks to the extraordinary
brightness of this maximal starburst \citepalias{canameras17b}. Since those lines were observed at the same time with the same array
configuration, they are perfectly suited to measure magnification factors, independent of beamsize variations. Figure~\ref{fig:almamaps}
shows that CO(4--3) and HNC(5--4) (blended with CN(4--3)) are nearly cospatial in the image plane, with only small spatial offsets
of 0.1--0.2\arcsec\ between the emission peaks. Line emission on the southern arc presents similar clumpiness in both
tracers. Finding similar distributions of both tracers along the partial Einstein ring already gives strong evidence that they
probe nearly the same regions in the source plane. As in \citetalias{canameras18b}, we used the amplification maps from our
best-fit lens model of PLCK\_G244.8+54.9 \citep{canameras17a} and the ALMA line-imaging in Fig.~\ref{fig:alma} to compute the
luminosity-weighted magnification factor of HNC(5--4). We found $\mu = 16.3 \pm 1.8$, lower than the factor $\mu = 22.0 \pm 1.3$
for CO(4--3) \citepalias{canameras18b}. This difference is higher than the 1$\,\sigma$ noise uncertainties and the $<$5\% systematics
expected from the strong lensing model \citepalias{canameras18b}, but remains comparable to other measurement uncertainties. The
different values might correspond to the small spatial offsets seen in Fig.~\ref{fig:almamaps}. Alternatively, $\mu_{\rm HNC}$ could
also be subject to other systematics from the line fitting (see Sect.~\ref{ssec:interfobs}) or from local variations of the CN(4--3)
line contamination, rather than differential magnification.

Moreover, the aforementioned FWHM and flux ratios of the blue and red line components in PLCK\_G244.8+54.9 provide further evidence that
differential lensing does not strongly affect these velocity components, which are likely to be tracing rotational motions in a fragmenting
gaseous disk \citepalias{canameras17b}. As shown previously, we measured similar line widths for these velocity components, over CO transitions
from $J=3$--2 to $J=10$--9 and HCN(5--4), which span four orders of magnitude in critical densities, $n_{\rm crit}$. This would not be the case
in the presence of strong differential magnification between CO-emitting gas and higher-density gas phases traced by HCO$^+$(5--4) and
HCN(5--4). In that scenario, variations in the line FWHMs would indicate that each tracer probes distinct regions in the source that sample
the velocity fields differently. On the contrary, our component separation procedure suggests that we probe intrinsic regions with similar
levels of turbulence with both emission lines. Although we cannot rule out different spatial distributions on galaxy-wide scales, this favors
a multi-phase ISM over regions most strongly magnified by gravitational lensing that dominate the observed line emission. Our finding of
dense-gas line FWHMs within 1$\,\sigma$ of CO in PLCK\_G092.5+42.9 and PLCK\_G145.2+50.9 (see Table~\ref{tab:linefit}) is also consistent
with this scenario. Together with small differences in our direct estimates of $\mu_{\rm HNC(5-4)}$ and $\mu_{\rm CO(4-3)}$, this leads us to
ignore the impact of differential lensing in PLCK\_G244.8+54.9, as well as in PLCK\_G092.5+42.9 and PLCK\_G145.2+50.9.

\section{Properties of the dense-gas reservoirs}
\label{sec:denseprop}

\subsection{Diagnostics from line ratios}
\label{ssec:lineratios}

In this section, we compare the line ratios involving high-density tracer molecules in the GEMS with samples of low- and high-redshift
sources in the literature.

\subsubsection{Ratios between dense gas and $^{12}$CO}

\begin{figure}
  \centering
  \includegraphics[width=0.45\textwidth]{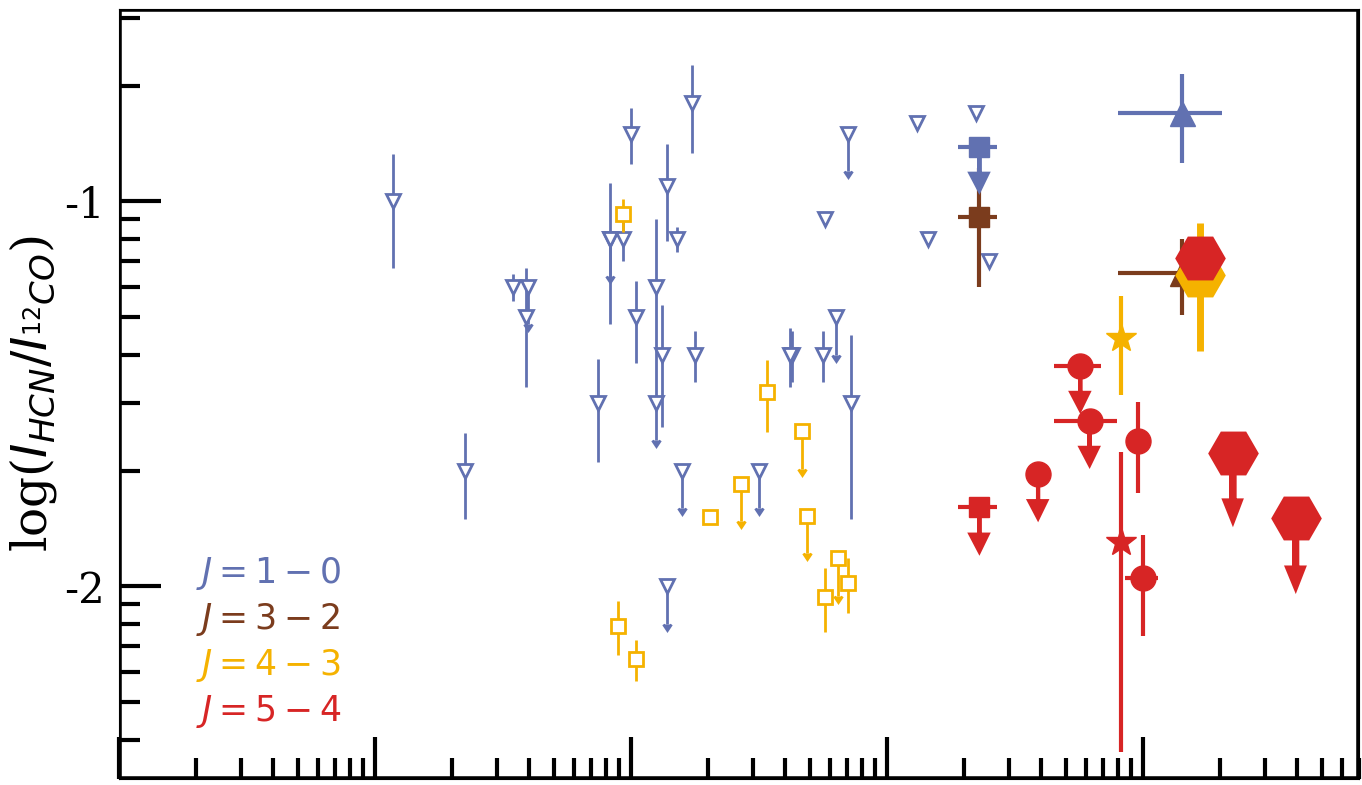}

  \includegraphics[width=0.45\textwidth]{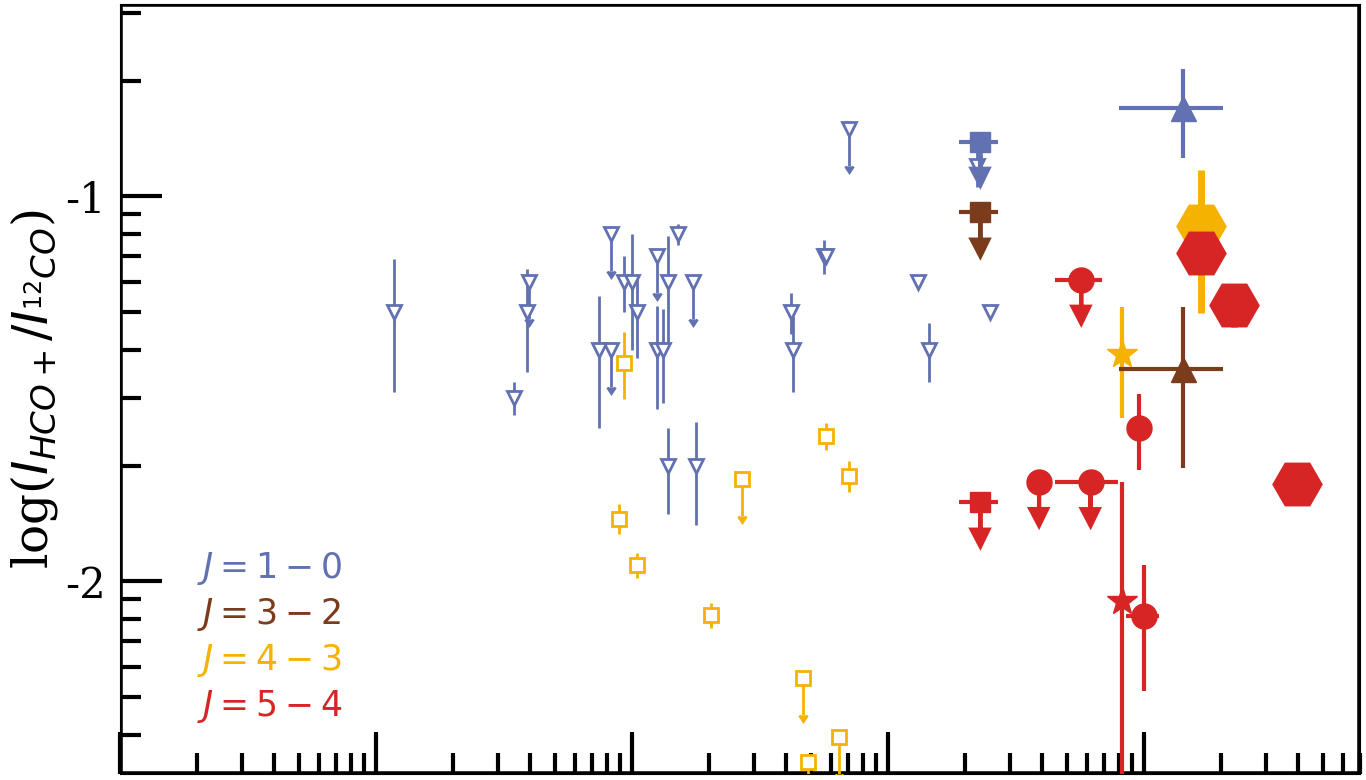}

  \includegraphics[width=0.45\textwidth]{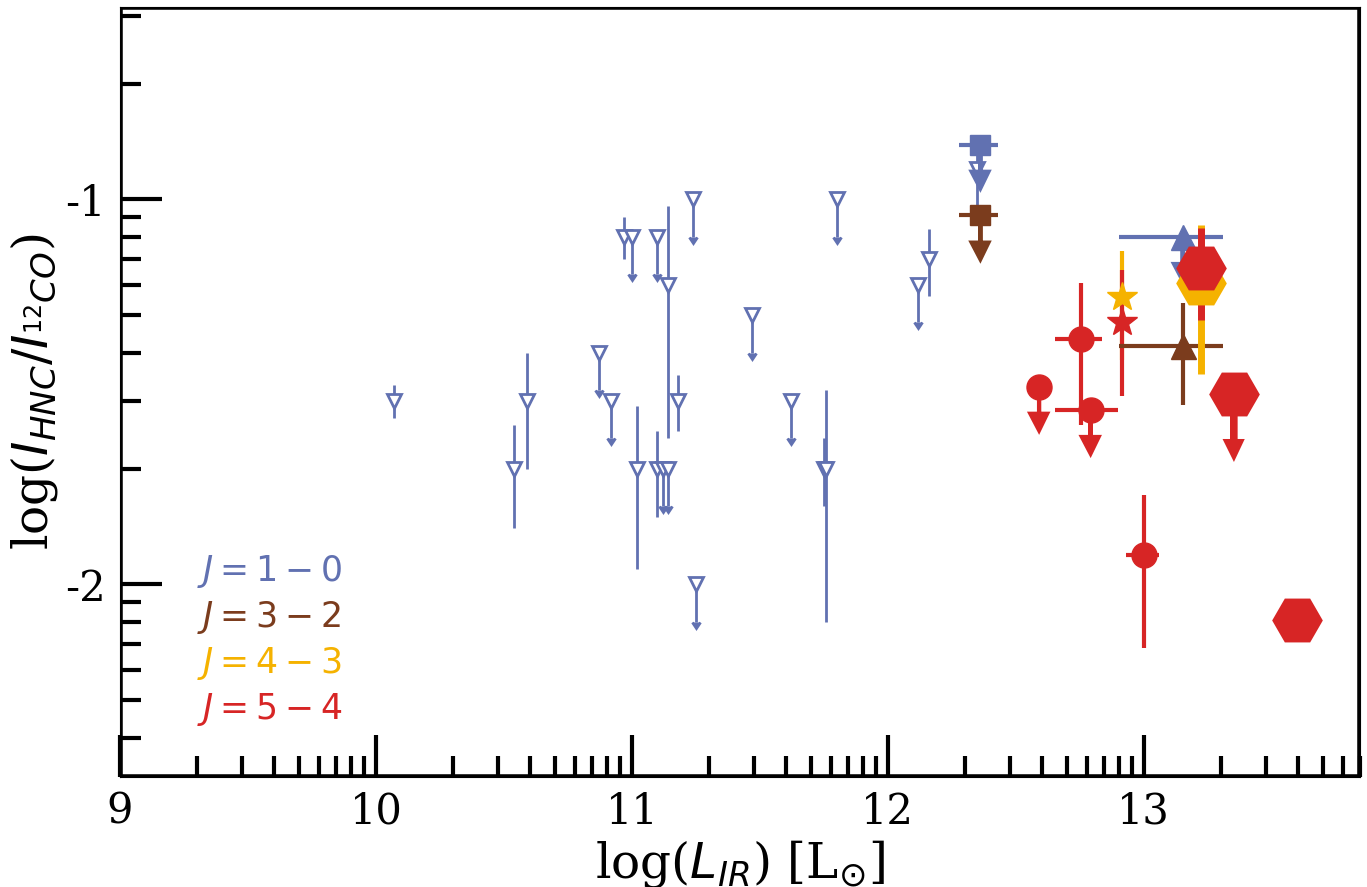}
  \caption{Line ratios between dense-gas tracers and CO as a function of the intrinsic $L_{\rm IR}$. Comparison samples are shown with the same
    symbols as in Fig.~\ref{fig:denseratios}. In particular, the line ratios of local LIRGs and ULIRGs are taken from \citet[][$J=1$--0
      transitions, blue triangles]{baan08}. The additional filled squares show HCN(3--2) in the Cosmic Eyelash and other upper flux density limits
    presented in \citet{danielson11}. The CO(4--3) fluxes of the local star-forming galaxies and ULIRGs of \citet[][yellow squares]{zhang14}
    are taken from \citet{rosenberg15}.}
  \label{fig:lineratios}
\end{figure}

We computed the flux ratio of dense gas to CO lines presented in \citetalias{canameras18b} using the same rotational level for both
transitions as is traditionally done in the literature. For sources without CO line measurements at the corresponding $J$ level, we used
extrapolated fluxes from the best-fit large velocity gradient (LVG) models from \citetalias{canameras18b}. Our measurements are listed
in Table~\ref{tab:lineratios} and plotted in Fig.~\ref{fig:lineratios}, together with values from the literature. Comparison samples
include nearby LIRGs and ULIRGs \citep{baan08,zhang14} and high-redshift lensed SMGs \citep{oteo17,bethermin18}, either in the ground or
mid-$J$ transitions. In particular, the line ratios of SPT SMGs from \citet{bethermin18} are given in terms of $J=5$--4, by converting
the CO(4--3) fluxes to CO(5--4) using the average ratio over this sample \citep[see the stacking analysis in][]{spilker14}.

When measured on the ground rotational level, ratios of high-density tracer molecules to CO are valuable proxies of the mass fraction of
dense gas over the bulk of the molecular gas reservoirs. Several studies in the past decade have characterized these ratios for populations
of active star-forming galaxies, LIRGs, and ULIRGs at low redshift, finding significant variations of the ratios HCN/CO, HCO$^+$/CO, and
HNC/CO, over a broad range from 0.01 to 0.15 \citep{graciacarpio08,baan08,costagliola11}. Ratios inferred for local normal spiral galaxies
are significantly lower \citep{jimenezdonaire19}, even when focusing on their inner regions, with galactocentric radii ${<}\,2\,$kpc, which
host more massive dense-gas reservoirs. On average, \citet{jimenezdonaire19} obtain HCN/CO $\simeq$ 0.025, HCO$^+$/CO $\simeq$ 0.018, and
HNC/CO $\simeq$ 0.011 over nine spiral disks, while \citet{baan08} find about 0.078, 0.052, and 0.043 for these ratios in LIRGs and ULIRGs.

These line ratios decrease gradually when using higher energy levels for both molecules \citep[e.g.,][]{juneau09}. For nearby LIRGs and
ULIRGs, the average HCN/CO $J=1$--0 ratios of \citet{baan08}, when both transitions are detected, are around 0.078, while the fluxes
from \citet{zhang14} and \citet{rosenberg15} lead to HCN/CO $\simeq$ 0.025 in $J=4$--3. PLCK\_G092.5+42.9 and PLCK\_G145.2+50.9 have
comparably low HCN/CO and HNC/CO ratios in the $J=5$--4 transition (Table~\ref{tab:lineratios}) and their dense-gas SLEDs likely peak
at lower $J$ levels than those of CO. However, for PLCK\_G244.8+54.9, the $J=4$--3 and $J=5$--4 ratios are consistent within 1$\,\sigma$,
suggesting very excited SLEDs in this extreme source. Variations in the line ratios demonstrate the need to carefully account for
the excitation of high-density tracer molecules in order to obtain a deep understanding of the gas phase directly fueling star
formation. We emphasize that since line emissions are not resolved in our dataset, these ratios are consistently probing source-averaged
quantities.

\begin{figure*}
  \centering
  \includegraphics[width=0.43\textwidth]{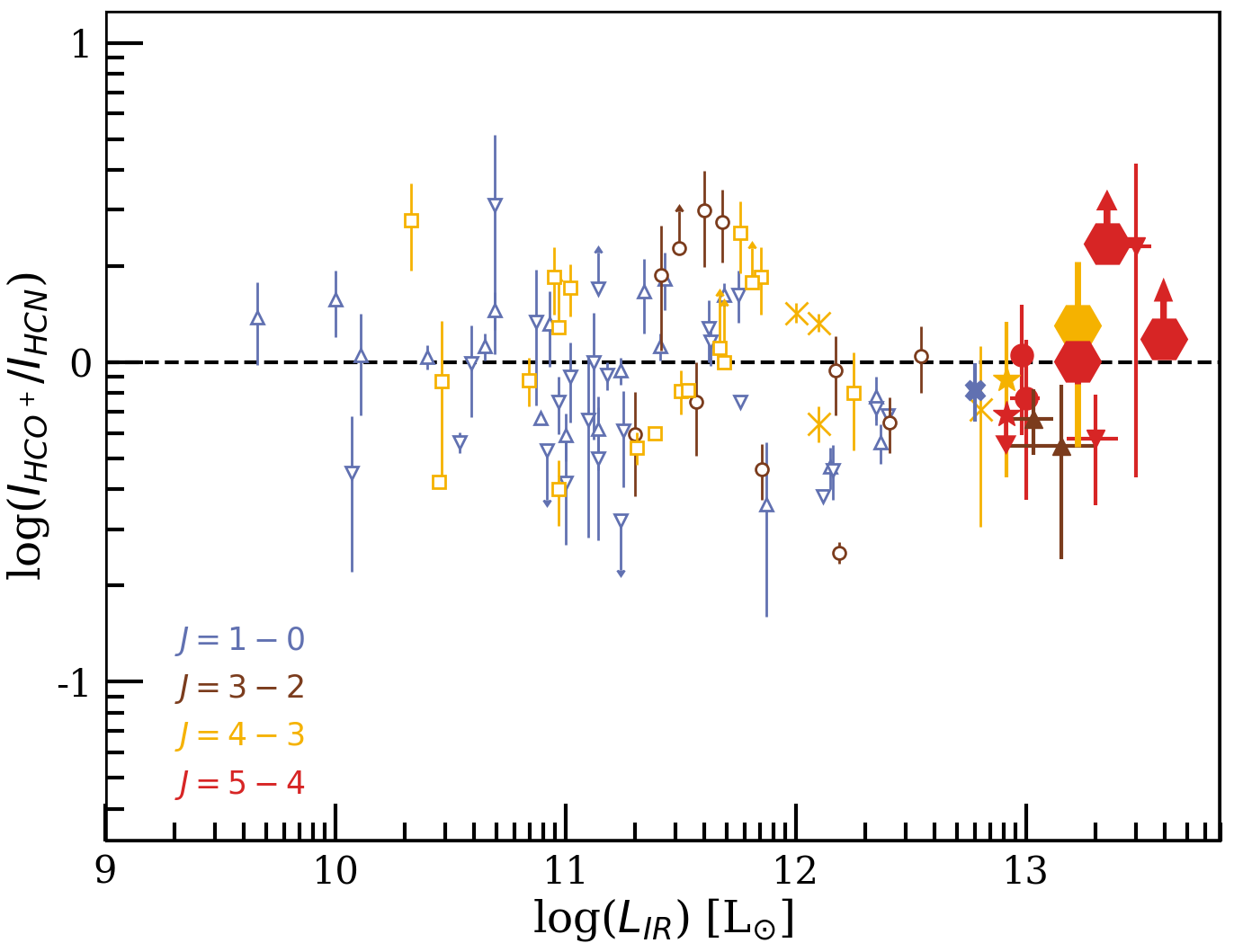}
  \hspace{1cm}\includegraphics[width=0.43\textwidth]{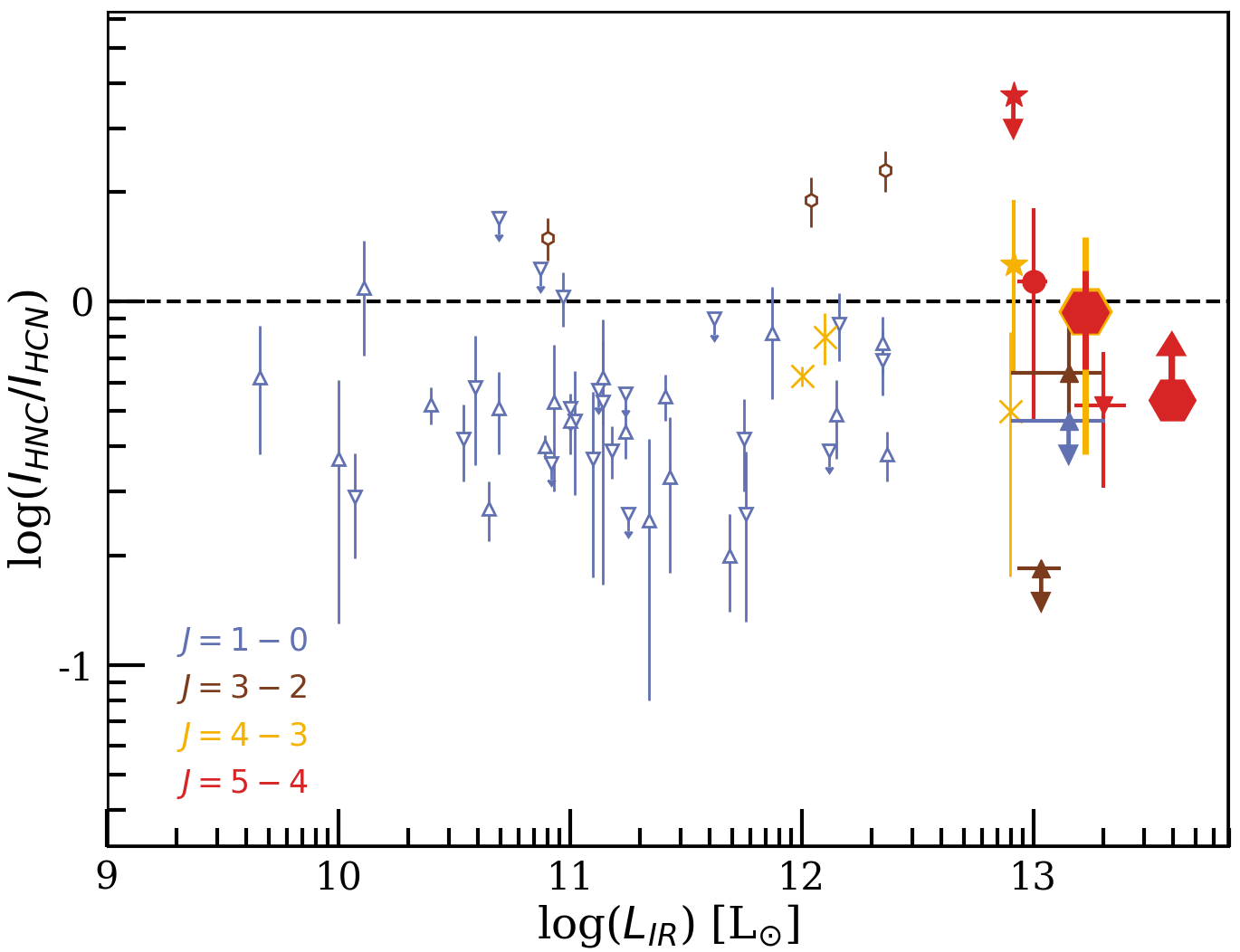}
  \caption{Line ratios of high-density tracer molecules measured in the GEMS as a function of the intrinsic IR luminosity. The quantity
    $L_{\rm IR}$ is integrated over the 8--1000~$\mu$m range and, for gravitationally lensed sources, error bars include uncertainties on the
    magnification factors. {\it Left:} HCO$^+$ to HCN flux ratios, with the GEMS shown as large hexagons. Symbols are color-coded according
    to the rotational levels of both transitions. The comparison samples include both low- ($z<0.1$, empty symbols) and high-redshift
    ($z\gtrsim1.5$, filled symbols) sources. We show the local samples of LIRGs and ULIRGs from \citet[][brown circles]{graciacarpio08},
    \citet[][blue upside-down triangles]{baan08}, and \citet[][blue upside-up triangles]{costagliola11}, as well as the normal
    star-forming galaxies and ULIRGs from \citet[][yellow squares]{zhang14}. Local ULIRGs with strong energy contributions from dust-obscured
    AGN of \citet{imanishi14} appear as yellow crosses. High-redshift line ratios include individual lensed SMGs from the H-ATLAS
    \citep[triangles,][]{oteo17,yang17b} and SPT \citep[circles,][]{bethermin18} surveys, the stack of SPT SMGs \citep[stars,][taking the
      average magnification factor over the sample]{spilker14}, and the Cloverleaf QSO \citep[blue cross,][]{riechers06}. {\it Right:} HNC
    to HCN flux ratios, using similar samples and symbols as above. In this panel, the brown hexagons show a few local starbursts from
    \citet{aalto07}.}
  \label{fig:denseratios}
\end{figure*}

Figure~\ref{fig:lineratios} shows that for each rotational level, line ratios in the three GEMS broadly agree with previous studies of
lensed SMGs. In general, HCN/CO and HCO$^+$/CO for $J_{\rm up}=4$ and 5 are slightly higher than the stack of SPT SMGs \citep{spilker14}.
Only upper limits for PLCK\_G092.5+42.9 and PLCK\_G145.2+50.9 approach the low HCN/CO(5--4) ratio of 0.013 in the stack. \citet{danielson11}
and \citet{oteo17} obtain higher HCN/CO and HCO$^+$/CO ratios in the Cosmic Eyelash, SDP9, and SDP11 than for the GEMS; this is not
surprising since these studies use $J=3$--2 and $J=1$--0 ratios, which, in low-redshift ULIRGs, are generally higher than for $J \geq 4$
due to the rapid decrease of the dense-gas SLEDs at $J \geq 3$ \citep{papadopoulos14}. The spatially-integrated HNC/CO ratio in
PLCK\_G244.8+54.9 is remarkably similar to the SPT stack, and it drops in PLCK\_G092.5+42.9 and PLCK\_G145.2+50.9 closer to the value of
0.01 in SPT0551$-$50 \citep[uncorrected for the CN(4--3) contribution,][]{bethermin18}. For each tracer, PLCK\_G145.2+50.9 exhibits the
lowest ratios among the GEMS.

In PLCK\_G244.8+54.9, the blue and red spectral components exhibit HCN/CO(5--4) ratios of 0.074$\pm$0.020 and 0.062$\pm$0.019, respectively,
using the CO line fitting from \citetalias{canameras18b}, which is consistent with the ratio of 0.071$\pm$0.009 obtained from integrated
line fluxes. These minor variations suggest similar distributions of gas phases traced by CO and HCN across the gaseous disk of this
maximal starburst. The velocity-integrated HCN(5--4) map of Fig.~\ref{fig:almamaps} broadly confirms that nearly 100\% of the dense-gas
emission arises from the southern arc, which also emits 65\% of the CO(4--3) emission. In addition, the HNC(5--4) (Fig.~\ref{fig:almamaps})
and CO(4--3) \citepalias{canameras17b} maps at a common resolution of 0.1\arcsec\ are well suited to compare both phases. The emission line
regions have very similar distributions, with 0.1--0.2\arcsec\ offsets between local peaks. Assuming a constant CN(4--3) line contamination,
the local variations of HNC/CO along the Einstein ring remain within a factor of two for all clumps identified in CO. The ratios are therefore
comparable over star-forming regions that are strongly magnified in PLCK\_G244.8+54.9, despite nearly four orders of magnitude difference
in critical density between the two transitions. Finally, we also find higher HCN/CO and HCO$^+$/CO integrated $J=4$--3 ratios in
PLCK\_G244.8+54.9 compared to the low-redshift galaxies from \citet{zhang14}, which could either indicate higher dense-gas fractions or
different excitations of the gas reservoirs probed by each molecule. We explore both hypotheses in Sects.~\ref{sec:denseprop} and
\ref{sec:sflaw} by using low-density ISM tracers to better constrain the dense-gas fraction and probe its impact on extreme star formation.

\subsubsection{Ratios between high-density tracer molecules}

\begin{table}
\caption{Line-flux ratios used as diagnostics.}
\centering
\begin{tabular}{llll}
\hline
\hline
\rule{0pt}{3ex} Source & $I_{\rm HCO^+/HCN}$ & $I_{\rm HNC/HCN}$ & $I_{\rm HCO^+/HNC}$ \\[+0.5em]
\hline
\noalign{\vskip 2pt}
PLCK\_G092.5+42.9 & $>$2.35 & $\oslash$ & $>$1.68 \\
PLCK\_G145.2+50.9 & $>$1.18 & $>$0.54 & 2.21 $\pm$ 1.19 \\
PLCK\_G244.8+54.9 & 1.30 $\pm$ 0.76$\dag$ & 0.94 $\pm$ 0.56$\dag$ & 1.38 $\pm$ 0.88$\dag$ \\
 & 1.00 $\pm$ 0.15 & 0.93 $\pm$ 0.28 & 1.08 $\pm$ 0.31 \\ 
\hline
\end{tabular}
\tablefoot{Flux ratios and 3$\sigma$ lower limits inferred from the line properties listed in Table~\ref{tab:linefit}. All rotational
  levels are $J=5$--4, except for PLCK\_G244.8+54.9 where ratios marked with a dagger correspond to $J=4$--3. The $\oslash$ symbols
  indicates that line fluxes are not available in either transition.}
\label{tab:denseratios}
\end{table}

\begin{figure*}
  \centering
  \includegraphics[width=0.31\textwidth]{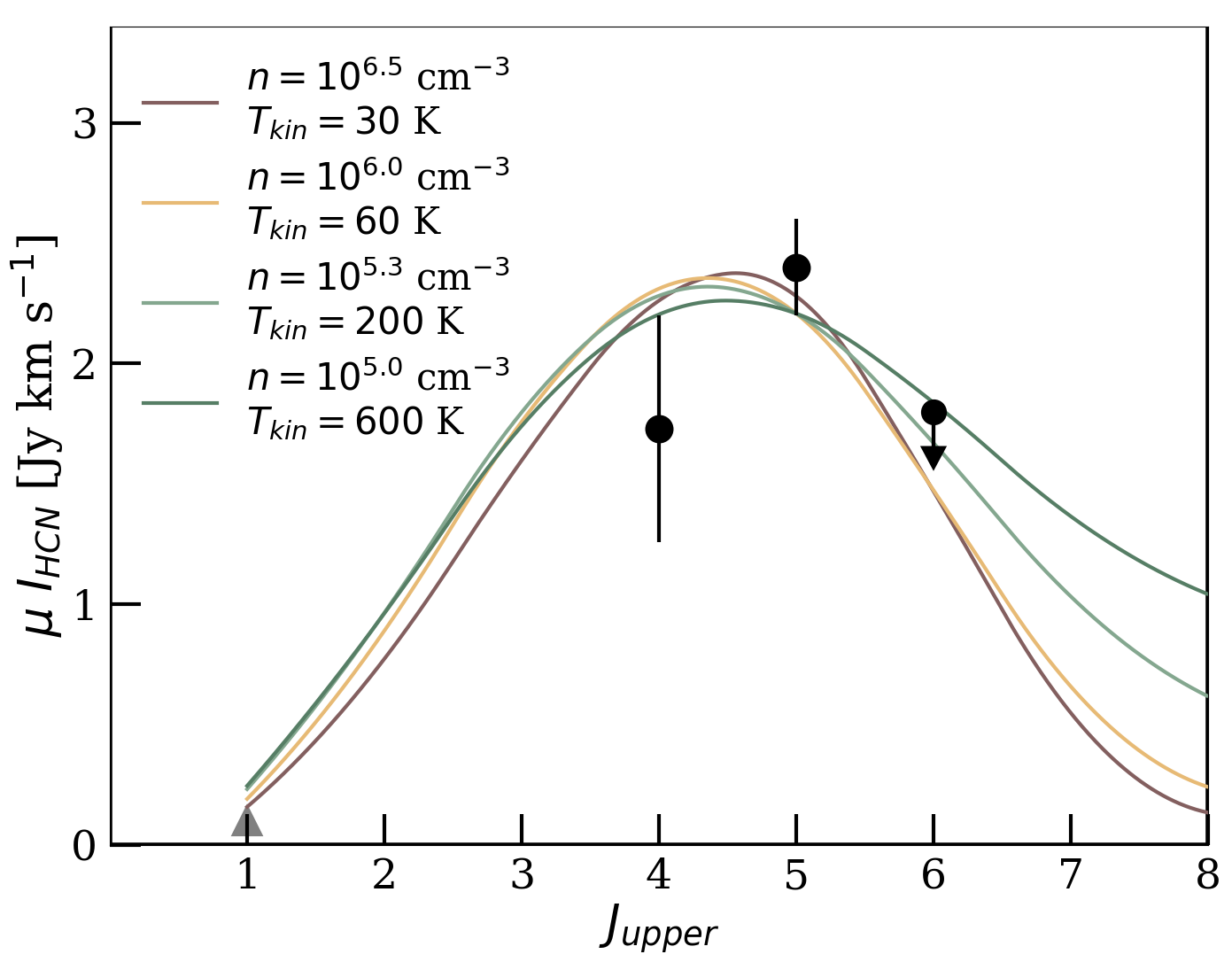}
  \includegraphics[width=0.31\textwidth]{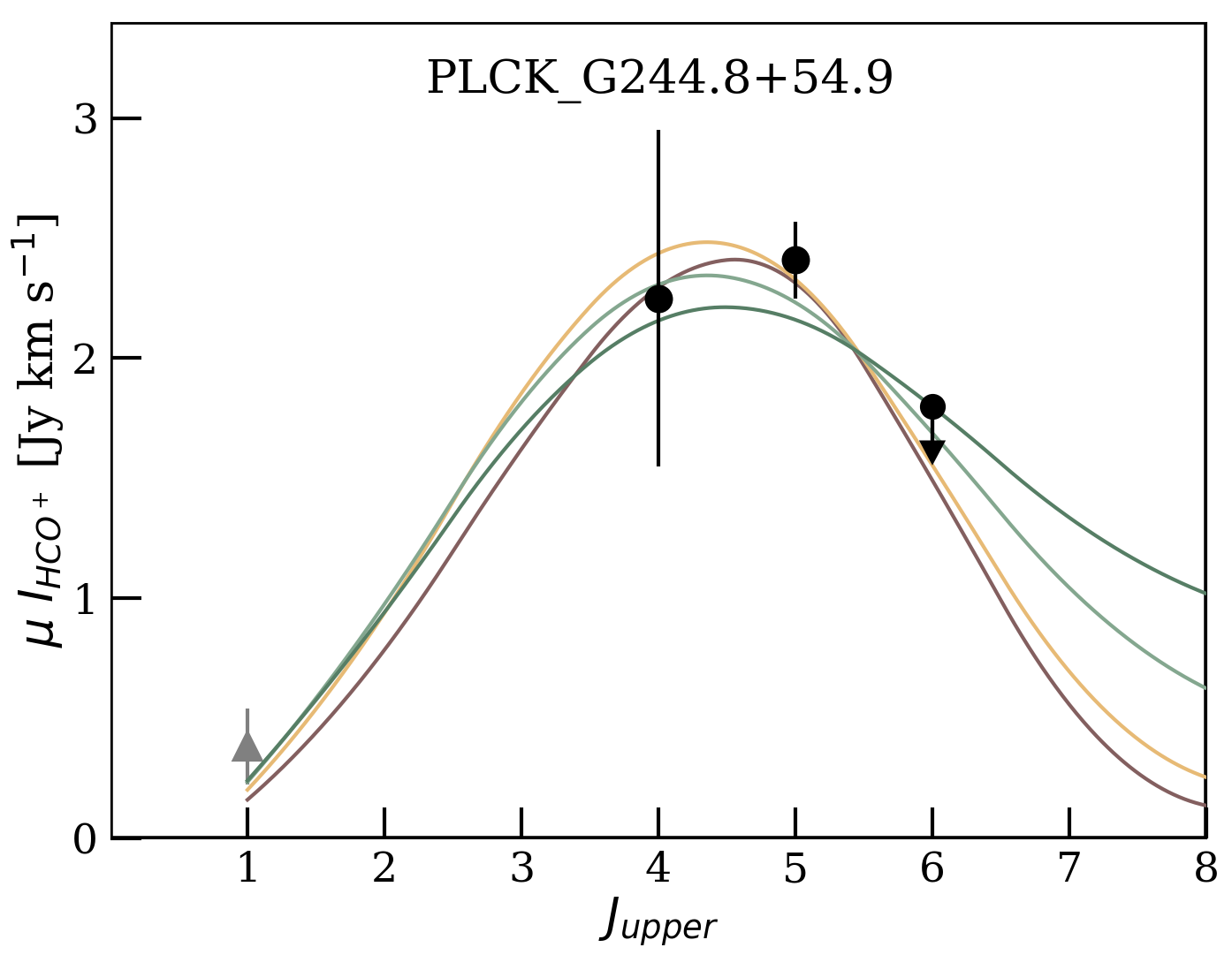}
  \includegraphics[width=0.31\textwidth]{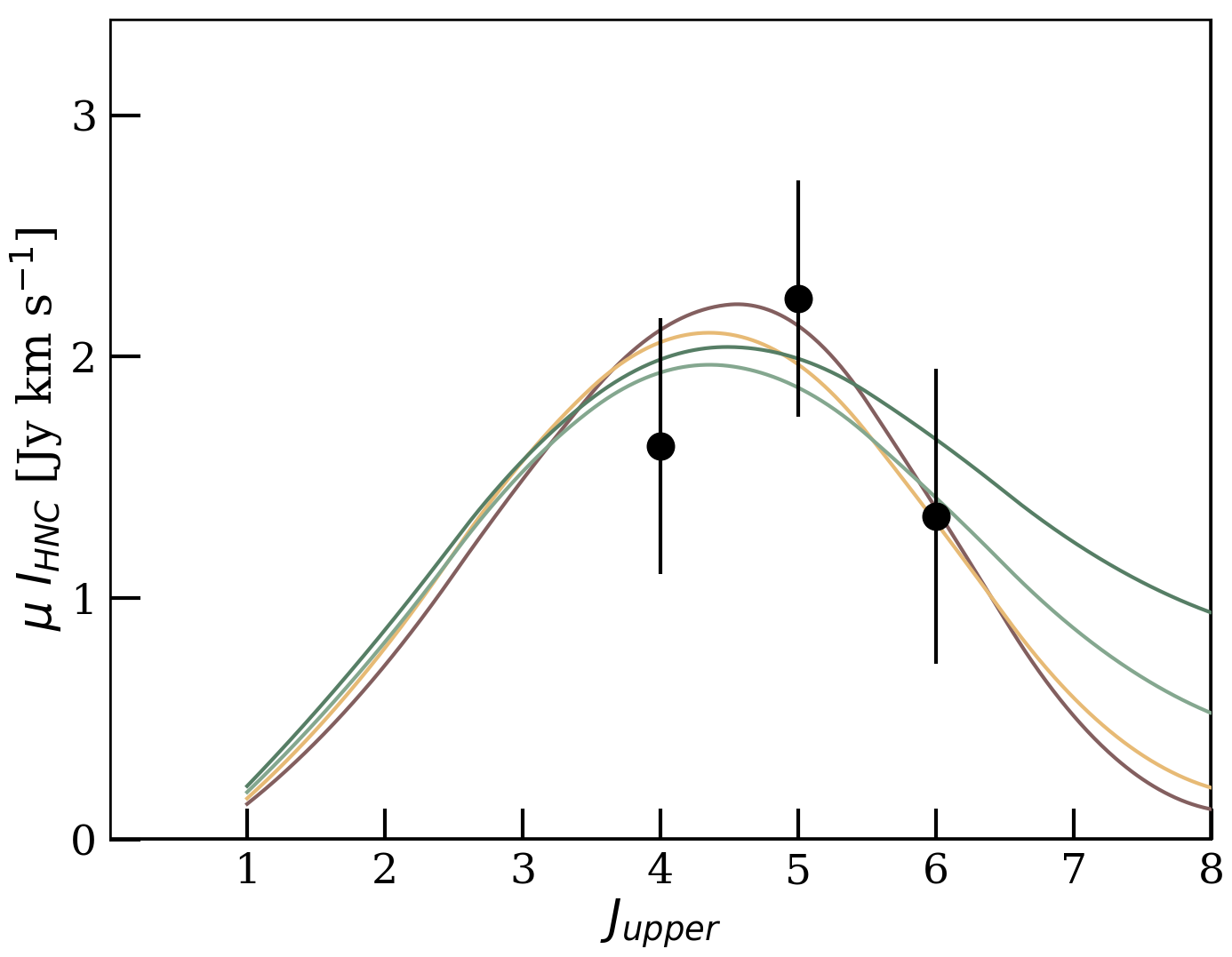} 
  \caption{HCN, HCO$^+$, and HNC SLEDs of PLCK\_G244.8+54.9 (black points). Gray triangles in the lower left of the first two panels refer to
    the dense-gas mid-$J$ to $J=1$--0 flux ratios in the local ULIRG Arp~220 \citep{imanishi07,greve09}, after normalizing to the HCN(4--3)
    and HCO$^+$(4--3) line fluxes in PLCK\_G244.8+54.9. Colored curves show a range of solutions to our collisional excitation models with
    {\tt RADEX}, combining the spatially-integrated intensity ratios of HCN, HCO$^+$, and HNC. These curves fit the three SLEDs reasonably
    well and illustrate the $n$--$T_{\rm kin}$ parameter degeneracies shown in more detail in Fig.~\ref{fig:lvgmod}. Lowering flux
    uncertainties and covering additional dense-gas transitions around the peak would improve the constraints.}
  \label{fig:sleds}
\end{figure*}

Table~\ref{tab:denseratios} and Fig.~\ref{fig:denseratios} illustrate the variations of HCO$^+$/HCN and HNC/HCN in the GEMS and other
samples of low- and high-redshift sources in the literature, covering four orders of magnitude in $L_{\rm IR}$. For PLCK\_G244.8+54.9,
$J=4$--3 and $J=5$--4 ratios are consistent within 1$\,\sigma$, as expected in the presence of collisional excitation, given the similar
critical densities of $J=4$--3 and $J=5$--4 transitions for both molecules. This similarity also suggests that our deblending of HNC(5--4)
and CN(4--3) in Sect.~\ref{sec:lineprop} is robust. Our comparison includes nearby gradually or more intensely star-forming galaxies from
different surveys that also include LIRGs and ULIRGs \citep{graciacarpio08,baan08,costagliola11,zhang14}, as well as high-redshift lensed
SMGs (with $L_{\rm IR}$ corrected for lensing magnifications), together with the Cloverleaf QSO \citep{riechers06}. These populations do
not show obvious systematic variations of the HCO$^+$/HCN line ratio for different rotational levels over the range $J_{\rm up}=1$--5
(color-coded in Fig.~\ref{fig:denseratios}), implying that both molecules have similar spatially-integrated SLEDs over a large variety
of active galaxies. Evidence remains scarce for the HNC/HCN ratio, which has been little studied for $J>1$ levels in the local Universe.
The ${\rm HNC/HCN \leq 1}$ ratio found by \citet{costagliola11} for the ground level in a sample of Seyferts, starbursts, and LIRGs, may
suggest significant differences in both SLEDs when compared with the ${\rm HNC/HCN > 1}$ ratio measured in Arp 220, Mrk 231, and NGC 4418 for
$J=3$--2 \citep{aalto07}. However, this would benefit from a more systematic study of HNC excitation for a larger sample.

Firstly, we measure an HCO$^+$/HCN(5--4) ratio of $1.00\pm0.15$ in PLCK\_G244.8+54.9, the only one of the GEMS with both lines detected,
and ${\rm HCO^+/HCN > 1}$ for PLCK\_G092.5+42.9 and PLCK\_G145.2+50.9, observed in $J=5$--4. These values are compatible with the regime of
luminous star-forming galaxies in the local Universe. They are generally higher than ${\rm HCO^+/HCN \lesssim 1}$ at $z>1.5$, although
PLCK\_G244.8+54.9 remains within 1$\,\sigma$ of some high-redshift sources in Fig.~\ref{fig:denseratios}. Lower limits are near the upper
regime of SMGs, and more comparable to the elevated ratio in G09v1.97 \citep{yang17b}, a strongly lensed major merger of two ULIRGs at
$z=3.634$, characterized at high angular resolution by \citet{yang19}. Secondly, we obtain similar brightness levels for HNC and HCN
lines in the GEMS. In PLCK\_G244.8+54.9 where both lines are detected, the HNC/HCN ratio is consistent with unity both for $J=4$--3 and
$J=5$--4, which is particularly close to the values for the SPT stack \citep{spilker14} and for SPT0551-50 \citep{bethermin18}. Other
high-redshift sources have lower ratios, down to the upper limit ${\rm HNC/HCN(3-2) < 0.2}$ in SDP11 \citep{oteo17}. HNC/HCN in
PLCK\_G244.8+54.9 is higher than average values over the sample of \citet{costagliola11} for $J=1$--0, while for PLCK\_G145.2+50.9, we
measure ${\rm HNC/HCN(5-4) > 0.54}$.

Both HCO$^+$/HCN and HNC/HCN line ratios in the GEMS are also higher than the average values of $0.7 \pm 0.2$ and $0.4 \pm 0.2$ obtained
for more gradually star-forming, spiral galaxies in the local Universe \citep[$J=1$--0 line detections,][]{jimenezdonaire19}. If there is
indeed no dependence on the $J$ level, together with the scaling relations in Sect.~\ref{ssec:lumrel}, this adds further evidence for lower
HCN line fluxes in the GEMS relative to local galaxies. In Sect.~\ref{ssec:heatingmecha}, we use these two ratios to probe the main heating
mechanisms within the densest molecular gas phase in the GEMS.

\subsection{Excitation of high-density tracers at high-redshift}
\label{ssec:sled}

Detailed analyses of high-density tracer molecule SLEDs are very rare, and most studies of the dense-gas ladder have focused on bright
local starbursts or (U)LIRGs \citep[e.g.,][]{papadopoulos14,saito18}. At high-redshift, the most detailed excitation analyses have been
conducted for the brightest QSOs, where HCN and HCO$^+$ rotational lines maintain exceptional fluxes up to transitions with upper levels
$J_{\rm up}=5$ to 6, due to the competition of radiative and collisional excitation mechanisms. For instance, \citet{riechers10} used the
$J=6$--5 transitions of HCN, HCO$^+$, and HNC in APM 08279+5255 at $z=3.91$, combined with HCN(5--4) and CO line fluxes \citep{weiss07}, to
highlight the major contribution of IR-pumping to the gas excitation in this strongly lensed quasar. The diagnostics remain hardly accessible
for SMGs at $z \simeq 2$--4, even for the brightest and most strongly magnified samples. \citet{spilker14} obtained ${>}\,3\,\sigma$ detections
for one or two different mid-$J$ levels by stacking ALMA spectra of 22 strongly lensed DSFGs from the SPT survey and found elevated gas
densities and temperatures using LVG models. In general, gas excitation for individual SMGs remains accessible only for the most extreme
lensed sources, such as the GEMS.

We derived radiative transfer models of HCN, HCO$^+$, and HNC line emission in PLCK\_G244.8+54.9, the only GEMS with multiple rotational
levels detected for these molecules. Figure~\ref{fig:sleds} shows the corresponding SLEDs, together with mid-$J$ to $J=1$--0 line ratios
in Arp~220 \citep{imanishi07,greve09}, an intense local starburst with comparable but slightly lower CO excitation \citepalias{canameras18b}.
The fluxes and upper limits in $J=6$--5 suggest that our dense-gas data cover the turnover in the SLEDs around $J=5$--4, at lower energy
levels than CO (which is at $J_{\rm up}\simeq6$ to 7, C18). This corresponds to higher dense-gas excitation than in the well-studied low-redshift
merger NGC~6240 \citep[][]{papadopoulos14}, one of the rare local starbursts with detailed analyses of the HCN and HCO$^+$ ladders peaking
at $J_{\rm up}\simeq3$ \citep[shock heating also plays a major role,][]{wang14}.

We inferred the average gas properties over star-forming cloud cores by assuming pure collisional excitation and ignoring other processes
(discussed in Sect.~\ref{ssec:heatingmecha}). Observed line ratios were compared with those predicted by the {\tt RADEX} code
\citep{vandertak07}, which solves the statistical equilibrium equations for a uniform medium. {\tt RADEX} is a non-LTE code using the LVG
method to compute the escape probability of emitted photons as a function of gas physical conditions and to predict the line intensities.
Constraints were exclusively taken from the $J_{\rm up}=4$ to 6 HCN, HCO$^+$, and HNC lines, which have comparable $n_{\rm crit}$ and can be
assumed to be cospatial.\footnote{In contrast, the $J_{\rm up}=6$ to 10 CO transitions have $\simeq 10^3$ times lower $n_{\rm crit}$ and can arise
  from warmer environments with different conditions \citep[e.g., the outer layers of giant molecular clouds, where HCN is
    dissociated,][]{boger05}. High-$J$ CO emission could also be influenced by mechanical feedback and not directly related
  to the star formation rate \citep{greve14}. Including these transitions in the excitation analysis would thus induce major uncertainties.}
We produced grids of line-intensity ratios with {\tt RADEX} for a range of molecular hydrogen number densities $n=10^3$--$10^7\,{\rm cm}^{-3}$
and gas kinetic temperatures $T_{\rm kin}=T_{\rm CMB}$ to $10^3\,$K, for an expanding sphere geometry, and using H$_2$ as the main collision partner
of high-density tracers with collision rates from \citet{flower99} and \citet{dumouchel10}. The column densities per unit velocity gradient
were fixed to $N_{\rm dense}/dv=10^{14}\,{\rm cm}^{-2}\,{\rm km}^{-1}\,{\rm s}$. This is consistent with our expectations that PLCK\_G244.8+54.9 is
highly obscured, with $N_{\rm H}>10^{24}\,{\rm cm}^{-2}$ akin to local ULIRGs such as Arp~220 \citep[e.g.,][]{wilson14}, and hosts HCN, HCO$^+$,
and HNC abundances relative to H$_{\rm 2}$ of 10$^{-9}$ to a few $\times10^{-8}$, comparable to local galaxies \citep{omont07} and Galactic
giant molecular clouds \citep[GMCs,][]{bergin96,helfer97}. This is also consistent with the linewidths in Table~\ref{tab:linefit}. Moreover,
these dense-gas column densities agree with the best-fit $N_{\rm CO}/dv$ values of 10$^{17.5}$--$10^{18}\,{\rm cm}^{-2}\,{\rm km}^{-1}\,{\rm s}$
in \citetalias{canameras18b} for [HCN/CO]$\,{\sim}\,$10$^{-4}$ \citep[e.g.,][]{omont07}. The models adopt a single gas excitation component
and a background radiation temperature of 10.9\,K, equal to $T_{\rm CMB}$ at $z=3.0$. We then computed the goodness-of-fit $\chi^2$ values
between the measured and predicted line ratios for each combination of $n$ and $T_{\rm kin}$.

\begin{figure}
  \centering
  \includegraphics[width=0.40\textwidth]{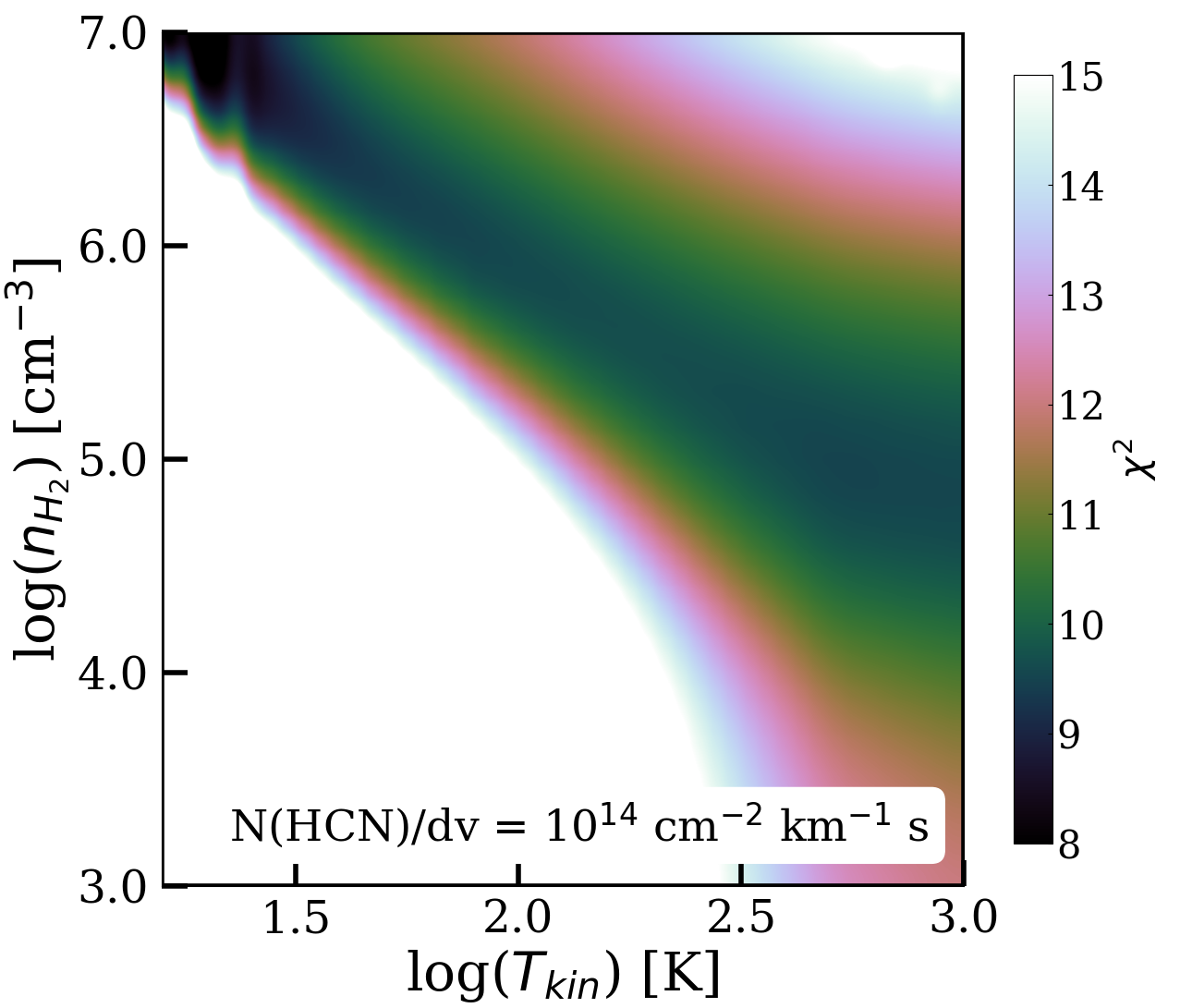}
  \caption{Results of collisional excitation models in PLCK\_G244.8+54.9 with {\tt RADEX} combining HCN, HCO$^+$, and HNC multi-$J$ line
    ratios. The color map shows the goodness-of-fit $\chi^2$ values (for four degrees of freedom), for each combination of $n$ and
    $T_{\rm kin}$ and illustrates the parameter degeneracies.}
  \label{fig:lvgmod}
\end{figure}

Figure~\ref{fig:lvgmod} shows the resulting $\chi^2$ distribution as a function of $n$ and $T_{\rm kin}$ and illustrates the strong parameter
degeneracies that prevent us measuring the exact gas conditions. The minimum $\chi^2 = 9.1$ (for four degrees of freedom) is obtained for
$n \simeq 10^{6.5}\,{\rm cm}^{-3}$ and $T_{\rm kin} \simeq 50\,$K, which would correspond to $T_{\rm kin} \sim T_{\rm dust}$ \citepalias{canameras15},
but similar fits are also obtained for solutions with $n > 10^{5.5}\,{\rm cm}^{-3}$ and $T_{\rm kin} \simeq 50$--100\,K, or with
$n < 10^{5.5}\,{\rm cm}^{-3}$ and $T_{\rm kin} \simeq 100$--600\,K. Figure~\ref{fig:sleds} shows the SLEDs for representative solutions,
and suggests that, even though the lines are moderately optically thick \citep[see, e.g., local ULIRGs,][]{scoville15}, lowering flux
uncertainties and covering additional transitions around and above the peak would significantly reduce parameter degeneracies.\footnote{More
  than adding $J=1$--0 line fluxes, if our simple modeling assumptions hold over the entire energy range. In any case, ground-level
  observations of HCN and HCO$^+$ remain essential to firmly constrain the mass fraction of molecular gas in the highest density phase.}
Our diagnostics combining all line ratios are in agreement with those for individual molecules, showing no evidence that HCN, HCO$^+$, or
HNC trace very different gas conditions \citep[in contrast to Arp~220,][]{imanishi10}. Overall, the models robustly exclude average
molecular gas densities of $n \ll 10^4\,{\rm cm}^{-3}$ and confirm that PLCK\_G244.8+54.9 hosts significant reservoirs of very dense gas
with $n \sim 10^5$--$10^6\,{\rm cm}^{-3}$. This is similar to the value $2 \times 10^5$~cm$^{-3}$ estimated in the dense cores of the Arp~220
nuclei \citep{scoville15}, as also shown in Fig.~\ref{fig:sleds} through the agreement between our LVG model extrapolations and line ratios
in Arp~220.

Interestingly, we obtain a common solution when adding the $J_{\rm up} \geq 7$ CO lines (dominated by the warm and compact CO-emitting phase
identified in \citetalias{canameras18b}) to the excitation analysis. This demonstrates that at least part of the high-excitation CO gas
emission in PLCK\_G244.8+54.9 arises from environments dense enough to be detected in HCN, HCO$^+$, and HNC. This could correspond to a
multiphase ISM, with one or multiple very dense ($n > 10^5\,{\rm cm}^{-3}$) and spatially-concentrated nuclear cores emitting HCN, HCO$^+$,
and HNC lines and hosting the Eddington-limited starburst identified in \citetalias{canameras17b}, which is distributed over an extended,
gas-rich and lower surface brightness disk-like component dominating the low- to mid-$J$ CO emission. Such ISM structures are common in
low-redshift ULIRGs and a similar scenario has been recently proposed by \citet{yang20} for a $z=3.6$ SMG with dense-gas conditions probed
by H$_{\rm 2}$O lines.

\subsection{Constraints on individual heating mechanisms}
\label{ssec:heatingmecha}

Non-collisional processes can enhance the HCN, HCO$^+$, and HNC line emissions in regions with moderate dense-gas fractions, and
vice versa, and can have important roles on the heating of gas deeply embedded within star-forming clouds.  This complicates our
interpretation of the dense-gas star-formation law. In this section, we determine whether intrinsic line ratios of HCN, HCO$^+$,
and HNC in the GEMS indicate an influence of such excitation mechanisms, from X-ray heating and IR-pumping \citep[e.g.][]{krips08},
to mechanical heating from gas outflows or supernova explosions \citep[e.g.,][]{loenen08}.

\subsubsection{Diagnostics from HCO$^+$/HCN}

Recent studies of nearby galaxies have highlighted that ${\rm HCO^+/HCN}$ is sensitive to the presence of XDRs, which are generally
produced by a nearby AGN rather than star formation. Radiative transfer models suggest that HCO$^+$ emission is enhanced in XDRs while
HCN remains unaffected by the elevated ionization rates in these environments, making the HCO$^+$/HCN $J=1$--0 and $J=4$--3 ratios higher
in XDRs than in PDRs \citep{meijerink05,meijerink07}. In NGC~1068, \citet{garciaburillo14} indeed measure a factor two increase of the
${\rm HCO^+/HCN}$ ratio close to the nucleus compared to the circumnuclear disk. In reality, these dense regions are nonetheless more
complex than simple XDRs and other processes certainly contribute \citep[e.g., mechanical heating through shocks,][]{tafalla10}.

Low-redshift studies show that spatially-integrated ${\rm HCO^+/HCN}$ ratios actually tend to be lower in AGN-dominated than in
star-formation dominated sources \citep[e.g.,][]{imanishi16,izumi16}, suggesting that the increase due to XDRs is rather a local
effect in the most extreme AGN environments. For instance, \citet{tan18} obtained lower ratios for galaxies with strong AGN contributions
in their sample of six nearby star-forming galaxies. Using $\lesssim500$-pc resolution ALMA line interferometry of nearby ULIRGs,
\citet{imanishi19} found a 15--20\% decrease of HCO$^+$/HCN $J=3$--2 ratios over compact regions located mainly in the central nuclear
cores, and systematically lower integrated HCO$^+$/HCN ratios $\lesssim$1 for those galaxies showing hints of strong dust-obscured AGN.
Generally, ${\rm HCO^+/HCN \lesssim 1}$ therefore seems to indicate AGN-dominated systems \citep[in agreement with ratios of][integrated
  over large radii]{garciaburillo14}.\footnote{On average, since a few individual nearby AGNs exhibit integrated ${\rm HCO^+/HCN}$ close
  to or even above unity \citep{imanishi14,imanishi16} as seen in Fig.~\ref{fig:denseratios}.}

In this context, HCO$^+$/HCN ratios and lower limits measured in PLCK\_G092.5+42.9 and PLCK\_G145.2+50.9 (see Table~\ref{tab:denseratios})
are consistent with star-formation dominated systems, as expected from the lack of a strong AGN contribution to dust heating (C15).
The ratios of PLCK\_G244.8+54.9 are more ambiguous as they lie just at the limit between the two regimes. While this also corroborates the
lack of XDR imprints on the high-to-low-$J$ CO line ratios in \citetalias{canameras18b}, we emphasize that an enhanced ${\rm HCO^+/HCN}$ is
not sufficient to firmly rule out the presence of XDRs \citep[see][]{privon15}. The elevated cosmic-ray fields expected from the intense
star-formation activity of the GEMS may also play an important role in setting the HCO$^+$/HCN ratios in dense cloud cores shielded from UV
fields, as argued for instance by \citet{schirm16} for the Antennae overlap region. The abundance of HCO$^+$ is most strongly affected by
variations in the cosmic-ray ionization rates and the resulting abundance of free electrons, but could be either enhanced or decreased
according to the literature \citep[e.g.,][]{papadopoulos07}. Given this ongoing debate and the possible influence of metallicity
\citep{braine17} or additional mechanisms \citep[see discussion in][]{oteo17}, the HCO$^+$/HCN ratio unfortunately cannot provide
definitive ISM diagnostics for the GEMS.

\begin{figure*}
  \centering
  \includegraphics[width=0.90\textwidth]{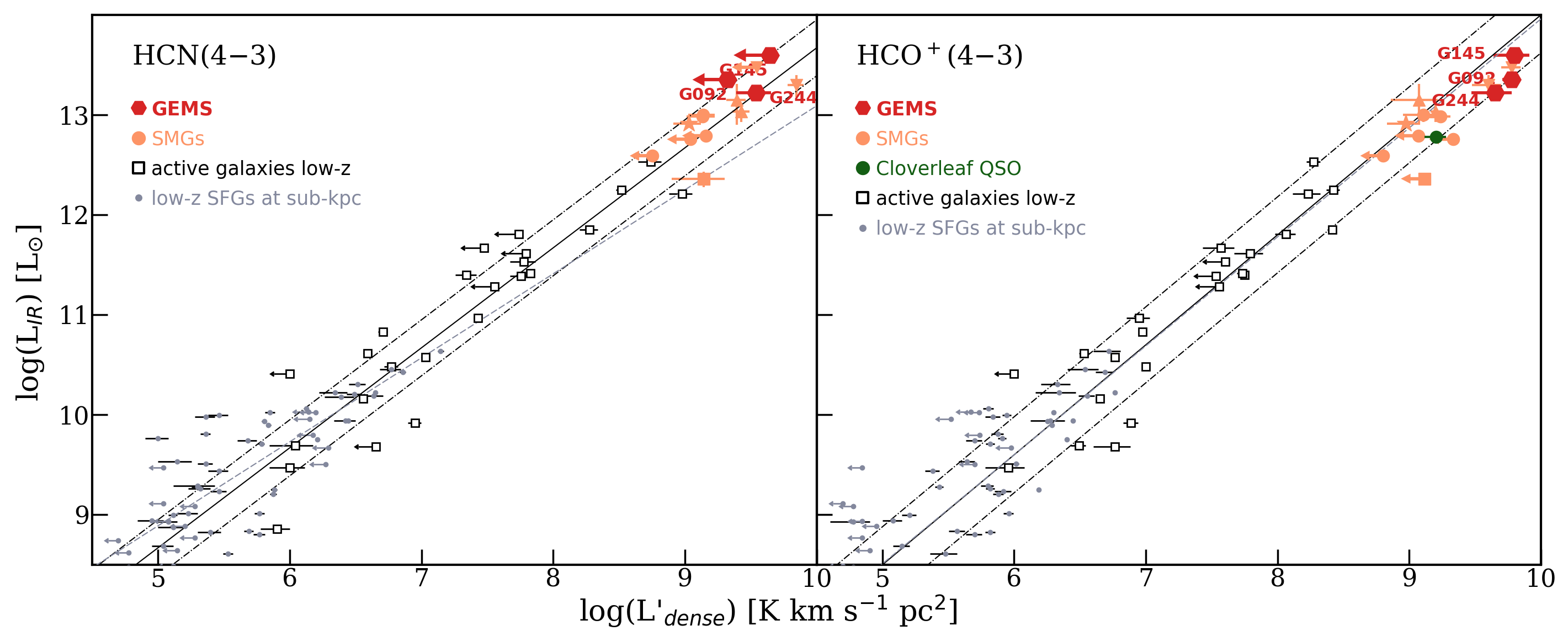}
  \caption{Scaling relations between the total infrared luminosities and the line luminosities of dense-gas tracers. The positions of the
    GEMS (red points) are compared with the low-redshift relations inferred for a diverse sample of normal blue star-forming galaxies,
    starbursts, and ULIRGs, and some AGN-dominated sources \citep[black squares,][]{zhang14}. Solid and dashed-dotted black lines show
    their best-fit relations and the $\pm1\,\sigma$ scatter, respectively. The relations obtained in \citet{tan18} for six nearby
    star-forming galaxies resolved at sub-kpc (gray points) appear as gray dotted lines. The HCN and HCO$^+$ luminosities of the GEMS
    are plotted for $J=4$--3, after correcting $L_{\rm IR}$ for $\mu_{\rm dust}$ and $L'_{\rm dense}$ for $\mu_{\rm gas}$, the magnification factors
    of the dust and gas components listed in C18. For PLCK\_G244.8+54.9, we plot the measured $J=4$--3 line luminosities, while for
    PLCK\_G092.5+42.9 and PLCK\_G145.2+50.9, we convert the $J=5$--4 line luminosities to $J=4$--3 using measurements in the circumnuclear
    disk near the Galactic center \citep[see text,][]{mills13}. Previous detections and upper limits of high-redshift sources are plotted
    as filled symbols and include dusty star-forming galaxies from a subset of the SPT sample \citep[orange circles,][]{bethermin18}, the
    stacking analysis of SPT sources \citep[orange stars,][]{spilker14}, four H-ATLAS lensed SMGs (SDP9 and SDP11 as orange triangles,
    \citealt{oteo17} and \citealt{yang17b}, NCv1.143 and G09v1.97 as upside-down triangles), and the Cosmic Eyelash \citep[orange
      squares,][]{danielson11}. We also show the Cloverleaf QSO \citep[green circle,][]{riechers11}. Measurements taken from the literature
    for $J=5$--4 or $J=3$--2 were converted to $J=4$--3 in the same way as for the GEMS, and strongly lensed sources were corrected for
    the magnification factors.}
  \label{fig:lumrel}
\end{figure*}

\subsubsection{Diagnostics from HNC/HCN}

The characterization of HNC/HCN provides additional constraints, in particular, regarding the identification of PDRs and XDRs.
Detailed analyses of Galactic molecular cloud cores acting as typical PDRs find ${\rm HNC/HCN} \simeq 0.1$--1.0 \citep{hirota98} and,
more recently, \citet{hacar20} obtained a similar range in Orion and highlighted that this line ratio strongly decreases for
higher gas kinetic temperatures. Interestingly, mid-$J$ HNC/HCN ratios are systematically below unity in classical PDR models
\citep{meijerink07}. \citet{loenen08} and \citet{costagliola11} proposed that additional mechanical heating from supernova
explosions might be needed to explain the lowest range, ${\rm HNC/HCN < 0.5}$. The lower limit ${\rm HNC/HCN > 0.54}$ in
PLCK\_G145.2+50.9 and the ratios in PLCK\_G244.8+54.9 are thus consistent with the regime where PDRs dominate (see
Fig.~\ref{fig:denseratios}). We compared these values with the PDR models of \citet{kazandjian15} that predict strong variations
in the presence of mechanical heating, especially for a highly enriched ISM and strong incident far-UV fields. We find that ratios
in the GEMS are close enough to unity to match expectations for classical PDR models without significant mechanical heating.
Additional transitions, ideally from the ground level, would be required to quantify the influence of this process.

To explain the ratio  ${\rm HNC/HCN \gtrsim 1}$ seen in Arp 220, NGC 4418, and Mrk 231, \citet{aalto07} invoked the influence of XDRs
with enhanced HNC abundance, or mechanisms such as IR-pumping affecting HNC excitation more efficiently than HCN\footnote{Because HNC
  can be pumped by continuum radiation at 21.5\,$\mu m$, making the process more efficient in environments with intermediate gas
  temperatures than for HCN, which is pumped by 14.0\,$\mu m$ photons.}. Following this interpretation, finding ${\rm HNC/HCN \simeq 1}$
in PLCK\_G244.8+54.9 for $J=4$--3 and $J=5$--4 does not necessarily indicate XDRs, a scenario that has already been disfavored on the
basis of HCO$^+$/HCN and high-to-low-$J$ CO line ratios in this source. Observations of the Galactic circumnuclear disk show that both
molecules can be excited by mid-IR radiation fields, but only HCN in star-forming clumps with elevated dust temperatures above 100\,K
\citep{mills13}. High-redshift DSFGs with $T_{\rm dust}=50\,$K, such as PLCK\_G244.8+54.9, therefore cover an intermediate regime where
only HNC transitions could be efficiently pumped, thereby increasing the overall HNC/HCN ratios. Similarly, \citet{riechers10} suggested
that radiative excitation from infrared pumping contributes to setting extreme ISM conditions in the quasar APM~08279+5255.

\section{Probing the dense-gas star-formation law}
\label{sec:sflaw}

\subsection{Scaling relations between dense-gas line and infrared luminosities}
\label{ssec:lumrel}

At low redshift, the star-formation rates of galaxies are tightly correlated with their total dense-gas content, with a power-law index
  of 1.0 \citep{gao04a}, lower than the index of 1.4 obtained from CO measurements \citep{kennicutt98}. The linear relation extends to
Galactic dense cores \citep{wu05,liu16}, down to $L_{\rm IR} \sim 10^3$--$10^6\,{\rm L}_{\odot}$ and to other high-density tracer molecules
\citep{zhang14}, suggesting that they are better proxies of the SFE than CO. Previous studies proposed this indicates that the star-formation
rates of low-redshift galaxies are driven by their dense-gas fractions, with a roughly constant SFE. The nature of the dense-gas
star-formation law nonetheless remains debated, with some studies finding shallow sub-linear relations \citep[e.g.,][]{bussmann08}. Further
diagnostics at high-redshift are particularly important to pursue the characterization of this $L_{\rm IR}$--$L'_{\rm dense}$ relation across
various galaxy classes.

Given the rarity of HCN and HCO$^+$ detections at high redshift, we place the GEMS on extrapolations of the scaling relations
inferred in the local Universe. These scaling relations are directly compared for $J=4$--3 to avoid relying on uncertain conversion
factors between mid-$J$ and ground rotational levels and to take direct advantage of the recent line surveys covering this
transition in low-redshift star-forming galaxies and ULIRGs. Due to the scarcity of active galaxies with detailed constraints
on the dense-gas SLEDs, we converted the $J=3$--2 or $J=5$--4 line luminosities of sources without $J=4$--3 observations, using
ratios measured in the Milky Way. \citet{mills13} present the most comprehensive excitation analysis of these molecules in the
Galactic central circumnuclear disk by fitting three transitions up to $J_{\rm up} = 9$ with non-LTE radiative transfer models. The
luminosities are taken as the average over the four regions across the gaseous disk modeled by \citet{mills13} and, for HCN, this
corresponds to $r_{\rm HCN,54}=L'_{\rm HCN(5-4)}/L'_{\rm HCN(4-3)} \simeq 0.78$ and $r_{\rm HCN,43}=L'_{\rm HCN(4-3)}/L'_{\rm HCN(3-2)} \simeq 0.86$.
We emphasize that uncertainties in applying these ratios to high-redshift environments are likely significant, but comparable to
alternative conversions, and that $r_{\rm HCN,54}$ is consistent with the value $0.91 \pm 0.33$ measured in PLCK\_G244.8+54.9. The line
luminosities in Table~\ref{tab:linefit} are derived following the method of \citet{solomon97}, and $L_{\rm IR}$ are integrated over
8--1000~$\mu$m in the rest-frame, from the best-fit mid-infrared to submillimetre composite template of each GEMS
\citepalias[see][]{canameras15}.

The high-redshift comparison sample is dominated by the bright strongly lensed sub-millimeter galaxies from SPT
\citep[][]{spilker14,bethermin18} and H-ATLAS surveys \citep[][]{oteo17,yang17b},\footnote{HCN and HCO$^+$ data for NCv1.143 and
  G09v1.97 are taken from C.~Yang's PhD thesis \citep{yang17b} available at {\tt https://tel.archives-ouvertes.fr/tel-01661478/document}}
in addition to the Cosmic Eyelash \citep[][]{danielson11}, and spans $1.5<z<4.0$. The published line luminosities were converted to
$L'_{\rm HCN(4-3)}$ and $L'_{\rm HCO+(4-3)}$, and corrected for the magnification factors listed in the corresponding papers (for the SPT
stack we used the average $\mu \simeq 12.5$ over the sample). In addition, we added the Cloverleaf at $z=2.56$, the only high-redshift
quasar with high-density tracer molecules detected for $J=4$--3 \citep{riechers11}. The GEMS fall on the upper envelope of luminosities
measured in these high-redshift sources, close to H-ATLAS strong lenses.

In Fig.~\ref{fig:lumrel}, we set PLCK\_G092.5+42.9, PLCK\_G145.2+ 50.9, and PLCK\_G244.8+54.9 (corrected for $\mu_{\rm CO}$ from
\citetalias{canameras18b}) relative to extrapolations of $L_{\rm IR}-L'_{\rm dense}$ relations for nearby galaxies. Firstly, the
GEMS are compared to relations from \citet{zhang14} drawn from $J=4$--3 line detections in the central regions of diverse nearby
galaxies, from normal blue star-forming galaxies to starbursts and ULIRGs, including a fraction of sources with significant AGN
contribution. Secondly, we juxtapose the scaling relations in the central regions of nearby star-forming galaxies at 0.2--1.0\,kpc
resolution from the MALATANG survey \citep[also for HCN(4--3) and HCO$^+$(4--3),][]{tan18}. For HCO$^+$, \citet{tan18} find good
agreement with lower resolution observations, both at low and high redshift, but for HCN, they find significantly lower, sublinear
slopes (see Fig.~\ref{fig:lumrel}).

The GEMS fall near the best-fitting relations from \citet{zhang14}, and PLCK\_G244.8+54.9 is fully consistent with their 1$\,\sigma$
scatter. However, upper limits of 2.1 and $4.3\times10^9\,{\rm K}\,{\rm km}\,{\rm s}^{-1}\,{\rm pc}^2$ for PLCK\_G092.5+42.9 and
PLCK\_G145.2+50.9, respectively, suggest a significant deficit of HCN(4--3) emission compared to expectations from the local Universe.
Interestingly, we found minor, but non-zero AGN contributions to dust heating for these two GEMS in \citetalias{canameras15}. Correcting
for this possible contribution and using purely star-formation-driven $L_{\rm IR}$ would bring these points closer to the low-redshift
relation. If not intrinsic, this HCN deficit in PLCK\_G092.5+42.9 and PLCK\_G145.2+50.9 could well be due to differences in $r_{\rm HCN,54}$ 
between Galactic environments and more extreme starbursts \citep[see, e.g.,][]{papadopoulos14}. The relation of \citet{tan18} better
reproduces the elevated HCN line luminosity in the Cosmic Eyelash \citep{danielson11}, but not the position of the GEMS, probably
because it is drawn from less active galactic regions covering a smaller range in $L_{\rm IR}$.

The three GEMS fall below the best-fit $L_{\rm IR}-L'_{\rm HCO+(4-3)}$ relation, near the two H-ATLAS sources of \citet{yang17b}. These
sources and other SMGs suggest there could be a trend towards lower values of log($L_{\rm IR}$)/log($L'_{\rm HCO+}$) for the brightest
high-redshift galaxies. Offsets for the GEMS even exceed the 1$\,\sigma$ scatter of the \citet{zhang14} relation when replacing
  the $L_{\rm IR}$ values in Fig.~\ref{fig:lumrel} by purely star-formation-driven $L_{\rm IR}$. This suggests that the GEMS favor
slightly lower slopes than $n=1.10 \pm 0.05$ from \citet{zhang14}. Although we could not rule out radiative excitation via IR pumping
or mechanical heating in PLCK\_G244+54.9 and PLCK\_G145.2+50.9 in Sect.~\ref{ssec:heatingmecha}, it is also possible that elevated
$L'_{\rm HCO+}$ traces variations in the degree of ionization of molecular gas at high-redshift. HCO$^+$ abundances depend
on the formation of H$_3^+$ and the abundance of free electrons in dense cloud cores, which are mainly affected by cosmic-ray ionization
\citep{papadopoulos07}. Enhanced fields should in principle increase the abundance of free electrons and reduce the abundance of HCO$^+$
molecules. Hence, Fig.~\ref{fig:lumrel} could possibly indicate lower cosmic-ray fields than expected in highly star-forming environments
as in the GEMS.

In summary, given the larger scatter of low-redshift relations and uncertainties in interpreting HCO$^+$ emission, the GEMS are consistent
with a unique dense-gas star-formation law, suggesting that their extreme star-formation rates are primarily driven by massive reservoirs
of dense gas. We did not attempt to characterize $L_{\rm IR}-L'_{\rm HNC}$ since this relation remains hardly constrained at low redshift and
it would be affected by uncertainties in deblending HNC(5--4) and CN(4--3) in the GEMS on local and global scales.

\subsection{The dense-gas fractions}
\label{ssec:fdense}
      
\begin{figure}
  \centering
  \includegraphics[width=0.48\textwidth]{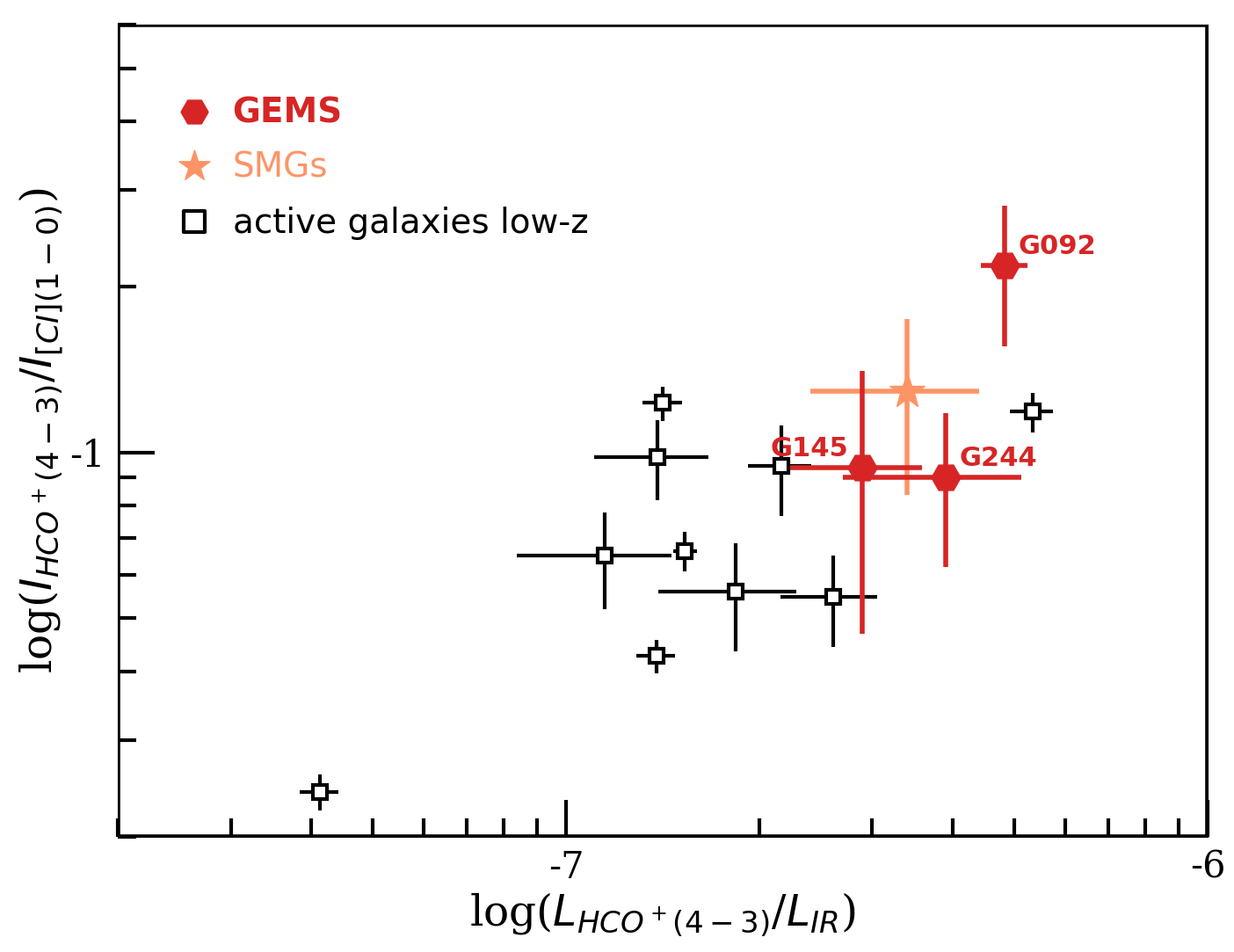}
  \caption{Flux ratio between HCO$^+$(4--3) and [\ion{C}{i}](1--0), a proxy of the global dense-gas fractions, as a function of
    $L_{\rm HCO+}/L_{\rm IR}$, tracing the depletion time of dense-gas reservoirs. The GEMS (red points) are compared with the local
    star-forming galaxies from \citet{zhang14} (black squares) and with the stack of SPT lensed SMGs (orange star).}
  \label{fig:fdense}
\end{figure}

An increase of the mass fraction of gas in the dense phase has been previously invoked to explain the outstanding star-formation
activities of high-redshift starbursts such as the GEMS \citep{daddi10b}. To study the evolution of this fraction at high redshift,
we use the ratios of mid-$J$ dense-gas emission lines from the current analysis to CO(1--0) from \citet{harrington18} or [\ion{C}{i}](1--0)
from \citetalias{nesvadba19}. Molecular gas masses derived from these two emission lines in PLCK\_G092.5+42.9, PLCK\_G145.2+50.9, and
PLCK\_G244.8+54.9 are roughly consistent with each other and suggest that these are two reliable tracers of the bulk of the gas reservoirs
in the GEMS \citepalias{canameras18b,nesvadba19}. Directly using the $J=4$--3 or $J=5$--4 emission lines of HCN and HCO$^+$ allows us to
compare their line ratios to CO(1--0) and [\ion{C}{i}](1--0) with those obtained in low-redshift active galaxies for the same transitions;
this should be independent of the conversion factors to the ground rotational levels, which are not well constrained by our LVG models
(see Sect.~\ref{ssec:sled} and the range of acceptable models in Fig~\ref{fig:sleds}). These modeling uncertainties and the subthermal
excitation of HCN and HCO$^+$ at mid-$J$ prevent us from estimating dense-gas masses from extrapolations to $J=1$--0, as usually done
for local ULIRGs \citep[e.g.,][]{gao04a}; however, we can still infer valuable diagnostics on the dense-gas fractions in the GEMS. 

The line-intensity ratios are summarized in Table~\ref{tab:lineratios}. For PLCK\_G244.8+54.9, we used the average [\ion{C}{i}](2--1) 
to [\ion{C}{i}](1--0) ratio of 0.86 in three other GEMS from \citetalias{nesvadba19} to predict the [\ion{C}{i}](1--0) line flux. For
PLCK\_G145.2+50.9, no CO(1--0) detection is available and we converted from mid-$J$ CO using the average ratio $I_{\rm CO(4-3)}/I_{\rm CO(1-0)}
\simeq 5.1$ in the GEMS from \citetalias{canameras18b}. The flux ratios listed in Table~\ref{tab:lineratios} cover the range 0.05--0.50
and do not list uncertainties when the [\ion{C}{i}] or CO line flux is converted from another transition. In
Fig.~\ref{fig:fdense}, we plot the HCO$^+$(4--3)/[\ion{C}{i}](1--0) ratios, which are available both for the three GEMS (after converting
from $J=5$--4 to $J=4$--3 for PLCK\_G092.5+42.9 and PLCK\_G145.2+50.9 as above), and for the large sample of low-redshift star-forming
galaxies and (U)LIRGs of \citet{zhang14}. This ratio is plotted as a function of $L_{\rm HCO+}/L_{\rm IR}$, a proxy of the depletion time of
the dense-gas reservoirs.

The line-intensity ratios in the GEMS fall near the upper range of values in local LIRGs, taken from the HCO$^+$(4--3) and [\ion{C}{i}](1--0)
line detections from \citet{zhang14} and \citet{rosenberg15}, and near the stack of SPT lensed SMGs from \citet{spilker14}. As expected,
such ratios between dense and total gas tracers are 1--2 orders of magnitude higher than in more gradually star-forming environments in
the local Universe \citep[see comparison with][in CO(1--0)]{tan18}. Provided that the GEMS and their most active low-redshift equivalents
have comparable HCO$^+$ excitations, as already suggested by our analysis of PLCK\_G244.8+54.9 in Sect.~\ref{ssec:sled}, and that
HCO$^+$/[\ion{C}{i}] is free of other systematics, obtaining similar ratios in Fig.~\ref{fig:fdense} rules out very different dense-gas
budgets in these two populations. Consequently, the dense-gas fraction in galaxies hosting the most violent starburst episodes
seems to have moderately evolved between $z \sim 3$ and $z \sim 0$, despite much higher global gas fractions and local gas-mass surface
densities at high redshift. Measurements of a few other high-redshift SMGs \citep{oteo17,bethermin18} have led to marginally higher
ratios, insufficient to establish a systematic difference. Finding important but not outstanding dense-gas reservoirs in the GEMS could
indicate they host GMC-like environments, over larger regions than the very compact nuclear cores of low-redshift galaxies \cite[e.g.,
  $\sim$10\%,][]{bemis19}. This scenario is entirely consistent with the similar spatial distributions of HNC(5--4) gas reservoirs and
less dense, CO-bright reservoirs in PLCK\_G244.8+54.9 (Fig.~\ref{fig:alma}). Further tests with a thorough analysis of local line ratios
over clump and intraclump regions in the GEMS would be beneficial, but are beyond the scope of this paper.

In the local Universe, the dense-gas depletion time shows little variation over a large range in $L_{\rm IR}$, for galaxies on global
\citep{gao04b} and sub-kpc scales \citep{tan18}, and for Milky Way GMCs \citep{wu05}. In theoretical models, a range of processes are
usually invoked for interpreting the positions of sources in the Schmidt-Kennicutt diagram, which do not require intrinsic variations
in the dense-gas depletion time \citep{krumholz12}. Moreover, concerning the GEMS, our high-resolution study of the star-formation law
and cloud stability in PLCK\_G244.8+54.9 \citepalias{canameras17b} did not show any evidence for an increase in the local SFE as traced by
CO. The lack of significant horizontal offset for the GEMS in Fig.~\ref{fig:fdense} is not, therefore, surprising
and further suggests that the SFE in the dense-gas phase remains roughly constant over cosmic history. These conclusions would remain
globally unchanged for HCN and CO(1--0), despite small differences in the relative positions of the three individual GEMS. Finally, we note
that Fig.~\ref{fig:fdense} is consistent with the lack of a significant correlation between the dense-gas fraction and the
dense-gas depletion time in local populations of star-forming galaxies \citep[][also for mid-$J$ HCN, HCO$^+$ levels]{tan18}.

\section{Conclusions}

In this paper, we use the high-density tracer molecules HCN, HCO$^+$, and HNC to characterize the ISM phase closely associated with
on-going star formation in three of the brightest dusty starburst galaxies at $z=3$--3.5, part of {\it Planck}'s Dusty GEMS sample. Our
detections of ten mid-$J$ lines in these sources with ALMA, NOEMA, and IRAM-30m/EMIR, together with additional upper flux limits,
significantly extend current diagnostics of dense molecular cloud cores at high redshift. 

With ALMA, we resolve HNC(5--4) line emission down to 0.1\arcsec\ in PLCK\_G244.8+54.9, which allows us to infer unprecedented constraints on
the spatial distribution of dense molecular gas at high redshift and to rule out a major impact of differential lensing between HNC and mid-$J$
CO in this extreme starburst. The similarity between these line morphologies in PLCK\_G244.8+54.9, and between spatially-integrated profiles
of HCN, HCO$^+$, HNC, [\ion{C}{i}], and mid-$J$ CO lines in the three sources suggests comparable distributions of very dense and more diffuse
gas reservoirs in $z \sim 3$ starbursts, at least over the regions that are most strongly magnified by gravitational lensing.

In two sources, PLCK\_G092.5+42.9 and PLCK\_G244.8+ 54.9, we obtain a good match between profiles of dense-gas and mid- to high-$J$ CO lines.
The lowest dense-gas line fluxes and the most significant differences in line profiles arise in the third source, PLCK\_G145.2+50.8, although
at lower S/N. This suggests that dense and more diffuse gas are better mixed in the first two GEMS than in PLCK\_G145.2+50.8, where our
line detections may probe fewer dense cores and more intraclump gas. 

The measured line ratios of ${\rm HCO^+/HCN \gtrsim 1}$ and ${\rm HNC/HCN \simeq 1}$ in the GEMS are akin to nearby ULIRGs and can be used
to characterize the gas-heating mechanisms. Despite unavoidable degeneracies in theoretical models, similar diagnostics in Milky Way GMCs
and nearby galaxies show consistency with PDRs in star-formation-dominated environments. The measured ratios do not require us to invoke
significant mechanical heating or AGN feedback in the GEMS. Moreover, the dense-gas to CO line ratios differ in our sample. These variations
highlight the importance of constraining the excitation of high density tracer molecules to improve our understanding on the gas phase
directly fueling star formation.

The extraordinary brightness of PLCK\_G244.8+54.9 offers a unique opportunity to characterize the HCN, HCO$^+$, and HNC excitation from
$J_{\rm up}=4$ to 6 lines covering the turnover in the SLEDs. Radiative transfer LVG models of integrated line emission show that these
transitions arise from a high-density phase with H$_{\rm 2}$ number densities $n \sim 10^5$--$10^6\,{\rm cm}^{-3}$, although important
degeneracies prevent us measuring the exact ISM conditions in these environments. Mid-$J$ to $J=1$--0 dense-gas excitation in
PLCK\_G244.8+54.9 is not very different from the intense local ULIRG Arp~220, which could indicate similar ISM structures and density
distributions, as already proposed for other high-redshift SMGs.

The three GEMS are consistent with extrapolations of dense-gas star-formation laws derived in the nearby Universe, adding further evidence
that the exceptional star-formation rates observed in the most active galaxies at $z \sim 3$ are a consequence of their massive reservoirs
of dense gas. This extends diagnostics of $L_{\rm IR}-L'_{\rm dense}$ relations beyond the luminosities of previous high-redshift samples.
Moreover, dense-gas-mass fractions in the GEMS inferred from mid-$J$ dense-gas emission lines to [\ion{C}{i}](1--0) are not exceptional, but
are close to other lensed SMGs and near the upper envelope of local ULIRGs. Hence, despite the higher global gas fractions and local gas-mass
surface densities observed at high redshift, the dense-gas budget of rapidly star-forming galaxies seems to have evolved little between
$z \sim 3$ and $z \sim 0$. Our results do not suggest important variations in the dense-gas depletion times of these populations, as also
predicted by theoretical models of star formation, and consistent with the local ISM conditions in PLCK\_G244.8+54.9.

This study highlights the need for pursuing investigations of the dense-gas properties and star-formation law near the peak of the
cosmic star-formation history. In particular, further observations of HCN, HCO$^+$, and HNC lines are needed to determine which molecule
best traces the overall dense-gas contents of DSFGs at high redshift and to derive precise diagnostics of the densities,
temperatures, and chemical abundances in cloud cores that are directly fueling star formation. Extending the coverage of dense-gas SLEDs
and probing larger samples will be highly valuable and achievable with ALMA for the brightest $z\sim2$--4 galaxies.

\section*{Acknowledgements}

We would like to thank the anonymous referee for comments that helped improve the paper. We thank A.~Omont for useful discussions and feedback
about this work, and the staff at the IRAM 30-m telescope for their excellent support during observations. C.Y. acknowledges support from an
ESO Fellowship. This paper is based on observations carried out under project number 108--14 with the IRAM 30-m telescope, and project S15CH
with the IRAM NOEMA Interferometer. IRAM is supported by INSU/CNRS (France), MPG (Germany), and IGN (Spain). This paper makes use of the
following ALMA data: ADS/JAO.ALMA\#2015.1.01518.S. ALMA is a partnership of ESO (representing its member states), NSF (USA) and NINS (Japan),
together with NRC (Canada), NSC and ASIAA (Taiwan), and KASI (Republic of Korea), in cooperation with the Republic of Chile. The Joint ALMA
Observatory is operated by ESO, AUI/NRAO and NAOJ.

\bibliographystyle{aa}
\bibliography{densegas}

\begin{appendix}

\section{Stacking analysis of EMIR spectra}

We stacked EMIR spectra of each individual HCN, HCO$^+$, and HNC transition in order to probe the average dense-gas properties and
excitation over the sample, at a level below what is possible for individual galaxies. The eight GEMS listed in Table~\ref{tab:obslog}
were included, with continuum-subtracted spectra either from the WILMA or FTS backends. We performed the stacking
following the method described in \citet{spilker14} and \citet{wilson17}. Spectra were shifted to a common redshift of $z=3.0$ according
to the best source redshift from CO \citepalias{canameras18b} because the dense-gas line redshifts are poorly constrained in most cases.
Although using common redshifts might be problematic in the presence of separate components with strong velocity offsets, this assumption
seems to hold true for dense-gas lines with centroids fitted independently (see Table~\ref{tab:linefit}). We then rescaled the flux densities
per spectral channel of each source to the values they would have at $z=3.0$ using Eq.~1 from \citet{spilker14}. For each transition, we
created a reference velocity grid covering $-1500$ to $+1500\,{\rm km}\,{\rm s}^{-1}$ and with $80\,{\rm km}\,{\rm s}^{-1}$ resolution. The
individual EMIR spectra were smoothed and interpolated to this new grid.

We first created mean stacks by computing the average flux density in each velocity bin. Between two and four GEMS were used in the stacks,
depending on the transition, with a uniform coverage over the complete velocity range. The resulting baseline rms varies from 1.5 to 2.5\,mJy
per $80\,{\rm km}\,{\rm s}^{-1}$ wide channel. We only obtained robust (${>}\,4\,\sigma$) detections of HCO$^+$(5--4) and HCO$^+$(6--5), with
fluxes of about 1.6 and $1.2\,{\rm Jy}\,{\rm km}\,{\rm s}^{-1}$, respectively. PLCK\_G092.5+42.9, PLCK\_G145.2+50.9, and
  PLCK\_G045.1+61.1 were used to stack HCO$^+$(5--4) and, for the $J=6$--5 transition, we combined PLCK\_G165.7+67.0 and PLCK\_G244.8+54.9.
Computing median stacks for the same transitions provided fluxes and upper limits consistent with those from the mean stacks, showing that
the previous results were not dominated by a few outliers. In addition, the average line properties over the sample are better described by
a weighted mean stack, where individual spectra are scaled to a common luminosity. We normalized the spectra to the average total infrared
luminosity of the GEMS using the source far-infrared luminosities $L_{\rm FIR}$ obtained from modified blackbody fits in \citetalias{canameras15}
(not corrected for lensing magnification), and created the weighted mean stacks. The results and number of detections are stable, except a
$\sim$30\% lower HCO$^+$(5--4) line flux than in the mean stack. Given the low fraction of lines detected in these three versions of the
stacks, we rather focus our analysis on line emission from the three brightest individual sources (PLCK\_G092.5+42.9, PLCK\_G145.2+50.9, and
PLCK\_G244.8+54.9).

\end{appendix}

\end{document}